\documentclass[prd,eqsecnum, 
notitlepage,
nofootinbib,superscriptaddress]{revtex4-1}
\global\arraycolsep=2pt
\usepackage{stmaryrd}
\usepackage{amsmath}
\usepackage{amssymb}
\usepackage{graphicx}
\usepackage{textcomp}
\usepackage{calrsfs}
\usepackage{yfonts}
\usepackage{bm}
\newcommand{\ben}{\begin{equation*}}
\newcommand{\een}{\end{equation*}}
\newcommand{\bean}{\begin{eqnarray*}}
\newcommand{\eean}{\end{eqnarray*}}

\newcommand{\nn}{\nonumber}
\newcommand{\be}{\begin{equation}}
\newcommand{\ee}{\end{equation}}
\newcommand{\bea}{\begin{eqnarray}}
\newcommand{\eea}{\end{eqnarray}}

\DeclareMathOperator{\Tr}{Tr}
\DeclareMathOperator{\csch}{csch}
\DeclareMathOperator{\Li}{Li}
\DeclareMathOperator{\Ei}{Ei}

\begin{document}

\title{Casimir Self-Entropy of a Spherical
Electromagnetic $\delta$-Function Shell}
\author{K. A. Milton}
  \email{kmilton@ou.edu}
  \affiliation{H. L. Dodge Department of Physics and Astronomy, 
University of Oklahoma, Norman, OK 73019 USA}

\author{Pushpa Kalauni}
  \email{pushpakalauni60@gmail.com}
  \affiliation{H. L. Dodge Department of Physics and Astronomy, University of 
Oklahoma, Norman, OK 73019 USA}

\author{Prachi Parashar}
  \email{prachi.parashar@ntnu.no}

\affiliation{Department of Energy and Process Engineering,
Norwegian University of Science and Technology, 7491 Trondheim, Norway}
  \affiliation{H. L. Dodge Department of Physics and Astronomy, University of 
Oklahoma, Norman, OK 73019 USA}
\author{Yang Li}
  \email{liyang@ou.edu}
  \affiliation{H. L. Dodge Department of Physics and Astronomy, University of 
Oklahoma, Norman, OK 73019 USA}

\begin{abstract}
In this paper we continue our program of computing Casimir self-entropies of 
idealized electrical bodies.  Here we consider an electromagnetic $\delta$-%
function sphere (``semitransparent sphere'') whose electric susceptibility
has a transverse polarization  with arbitrary
strength.  Dispersion is incorporated by a plasma-like model.
In the strong coupling limit, a perfectly conducting spherical
shell is realized.  We compute the entropy for both low and high temperatures.
The TE self-entropy is negative as expected, but the TM self-entropy requires
ultraviolet and infrared subtractions, and, surprisingly, is only positive
for sufficiently strong coupling.  Results are robust under different
regularization schemes.
\end{abstract}
\date\today
\maketitle

\section{Introduction}
\label{sec:intro}
The usual expectation, based on the notion that entropy is a measure of
 disorder, is that entropy should be positive.  However, there are 
circumstances in which entropy can take on negative values.  For example,
negative entropy is often discussed in connection with biological systems
\cite{schrodinger}.  More interesting physically is the occurrence of negative
entropy in black-hole and 
cosmological physics \cite{Cvetic:2001bk,Nojiri:2004pf}.

In Casimir physics, perhaps the first appearance of negative entropy occurred
in connection with the description of the quantum vacuum interaction between
parallel conducting plates.  If dissipation is present, 
the entropy of the interaction is positive at large distances, $aT\gg 1$, where
$a$ is the separation between the plates and $T$ is the temperature, 
but turns negative for short
distances.  Considered as a function of temperature, the sign of the entropy
changes as the temperature decreases, but does tend to zero as the temperature
tends to zero, in accordance with the Nernst heat theorem \cite{njp}.
Although perhaps surprising, this was not thought to be a problem because
this phenomenon only referred to the interaction part of the free energy, and
the total entropy of the system was expected to be positive.  Somewhat later
it was discovered that negative Casimir entropies also occurred purely
geometrically, for example between a perfectly conducting sphere and a 
perfectly conducting plane without dissipation \cite{most1,canaguier-prl,
canaguier-pra}, or between two spheres \cite{rodriguez-prb,rodriguez-qfext}.
When the distance times the temperature (in natural units) is of order unity,
typically a negative entropy region was present.  Since the effect was dominant
in the dipole approximation, this led to a systematic study of the phenomenon 
of negative entropy arising between polarizable particles, characterized by 
electric and
magnetic polarizabilities, or between such particles and a conducting plate.
For appropriate choices of these polarizabilities, these nanoparticles behaved 
like
small conducting spheres. We found that sometimes 
the entropy started off negatively for small $aT$, 
before eventually turning positive, and sometimes 
the entropy was first positive, turned negative for a while, and then turned 
positive again as $aT$ increased \cite{milton,ingold}.  
The combined effects of both geometry and dissipation are considered in
Refs.~\cite{umrath,bp}.

The occurrence of negative entropy, geometrically induced,
 sharpened the puzzle.  Again, the suspicion was that the self-entropies
of the bodies were much larger, and positive, yielding positive entropies 
always for the whole system.  This was borne out to some extent in the case of 
perfect conducting spheres.  There it turned out that although the self-entropy
of a conducting plate vanishes, the self-entropy of a conducting sphere is 
positive and is such that it precisely cancels the most negative interaction 
entropy between a sphere and a plate \cite{ly,fort}. 
More specifically, the two 
electromagnetic mode contributions to the entropy, the TE and TM terms, had 
opposite signs:  As expected \cite{hoye}, the 
TE was always negative, and the TM positive, the latter dominating the former.

In this paper, we carry the sphere self-entropy problem much further.  We 
consider a simple model of an electromagnetically coupled sphere, 
represented by
a $\delta$-function shell, with arbitrary coupling $\lambda$.  In the limit as 
the coupling tends to
infinity, this precisely corresponds to a perfectly conducting sphere.  This 
model generalizes the previously described electromagnetic $\delta$-function 
plate
\cite{Parashar:2012it}, and our electromagnetic $\delta$-function sphere 
\cite{prachi}, considered at zero temperature.
(The closely-related plasma spherical shell was considered earlier in
Refs.~\cite{bk,bartonps}.)
  As in our previous works,
we model the dispersive property of the shell by a plasma model.
  We discover that the finite-temperature
problem is much more complex than might have been anticipated.  Although it is
generally supposed that the divergences in the 
self-free energy are confined to the 
zero-temperature contribution, this is not the case:  The TM contribution to 
the entropy has both infrared and ultraviolet divergences, which violate the 
Nernst heat theorem, and hence require some sort of subtraction or 
``renormalization.''  When this is done, the TM self-entropy is no longer 
always positive.  It is positive only for sufficiently strong coupling, while 
the TE self-entropy is always negative as expected. So we have encountered  
new phenomena that will require further work to understand.

The outline of this paper is as follows.  In Sec.~\ref{sec3} we set up the 
formalism and obtain the general expressions for the free energy of the 
$\delta$-function sphere. The expressions are regulated by point-splitting
in time and in the angle on the sphere.  We also model the dispersive
properties of the shell with a plasma model, characterized by a dimensionless
coupling $\lambda_0$.
In Sec.~\ref{sec4} we consider the strong coupling limit, that of a perfectly 
conducting spherical shell.  
Because of the appearance of an infrared singularity, a 
renormalization of the temperature-dependent part of the
free energy is required.  After 
a temperature-dependent infrared-sensitive term is removed, the high
and low temperature results of Balian and Duplantier \cite{bd} are recovered.
Finite coupling behaviors are studied first in Sec.~\ref{sec5} using the 
uniform asymptotic expansion for the spherical Bessel functions:  At low 
temperature, this approach only yields the zero-temperature structure with 
divergences as seen in
Ref.~\cite{prachi}.  The low-order contributions to the entropy can actually be
found exactly in Sec.~\ref{sec:exactlow}, but again
the $O(\lambda_0^2)$ TE and the $O(\lambda_0)$   TM part exhibits 
divergences which violate the Nernst theorem and must be ``renormalized'' away.
The low temperature behavior is studied for both the TE and TM modes in 
Sec.~\ref{sec:lowt}.  The TE
low-temperature  entropy has a simple functional dependence on the coupling, 
while the TM mode again requires infrared renormalization, and exhibits a 
dependence on the coupling which is nonmonotonic, being positive for strong 
coupling, but changing
to negative as the coupling gets weaker.  Finally, the high-temperature limit 
is discussed in Sec.~\ref{sec:hit},
now using analytic regulation and the Chowla-Selberg formula.  
It is seen that the order of limits is 
important; we consider both the limits $aT\gg\lambda_0\gg1$, and 
$\lambda_0\gg aT\gg1$, where only 
the latter corresponds to the perfect conductor.  The TM mode again requires a
temperature dependent renormalization. 
The results of this paper are summarized in Sec.~\ref{sec7}.
 In the last  section of the paper we offer some concluding remarks.

We use natural units $\hbar=c=k_B=1$, and Heaviside-Lorentz electromagnetic 
units.

\section{Electromagnetic $\delta$-function plate}
\label{sec3}
As in Ref.~\cite{ly}, we can express the Casimir self-free energy of an object
with permittivity $\bm{\varepsilon}=\bm{1}+\bm{V}$ in symbolic form
\be
F=\frac{T}2\sum_{n=-\infty}^\infty \Tr\ln(\bm{1}-\bm{\Gamma}_0\mathbf{V}),
\label{fe}
\ee
expressed as a sum over Matsubara frequencies $\zeta_n=2\pi n T$, 
where the trace is over spatial coordinates and internal variables (tensor
indices).
Here $\bm{\Gamma}_0$ is the free electromagnetic Green's dyadic,
which satisfies
\be
\bigg[-\frac1{\zeta^2_n}\bm{\nabla}\times\bm{\nabla}\times-\bm{1}
\bigg]\bm{\Gamma}_0(\mathbf{r-r'})=\bm{1}\delta(\mathbf{r}-\mathbf{r}^\prime).
\ee
  It is
 convenient to define a divergence-free Green's dyadic, which
differs from this by a $\delta$-function term \cite{mds}:
\be
\bm{\Gamma}_0'(\mathbf{r-r'})=\bm{\Gamma}_0(\mathbf{r-r'})+\bm{1}\delta(\mathbf
{r-r'}), \quad \bm{\nabla}\cdot\bm{\Gamma}_0'=0.
\ee
This dyadic can be resolved in terms of vector spherical harmonics
\be
\mathbf{X}_{lm}=\frac1{\sqrt{l(l+1)}}\mathbf{L}Y_{lm}(\Omega),\quad
\mathbf{L}=\mathbf{r}\times \frac1i \bm{\nabla},
\ee
as follows:
\be
\bm{\Gamma}_0'(\mathbf{r-r'})=
\sum_{nlm}\bigg[-\zeta_n^2 g^0_l(r,r')\mathbf{X}_{lm}(\Omega)
\mathbf{X}^*_{lm}(\Omega')-\bm{\nabla}\times g^0_l(r,r')\mathbf{X}_{lm}(\Omega)
\mathbf{X}^*_{lm}(\Omega')\times\overleftarrow{\bm{\nabla}}'\bigg],
\ee
where in spherical coordinates $\mathbf{r}=(r,\Omega)=(r,\theta,\phi).$

For the case of a sphere of radius $a$ described by a
 semitransparent potential, $\mathbf{V}=\lambda\delta(r-
a)(\bm{1}-\mathbf{\hat r\hat r})$,
the potential is tangent to the surface of the sphere.
(This tranversality is required by Maxwell's equations \cite{Parashar:2012it}.)
The trace 
is then worked out by using the following orthonormality properties of
the vector spherical harmonics \cite{CE}:
\begin{subequations}
\bea
\int d\Omega \mathbf{X}_{l'm'}^*(\Omega)\mathbf{X}_{lm}(\Omega)&=&\delta_{ll'}
\delta_{mm'},\\
\int d\Omega f(r')\mathbf{X}_{l'm'}^*(\Omega)\bm{\nabla}\times g(r)
\mathbf{X}_{lm}(\Omega)&=&0,\\
\int d\Omega \bm{\nabla}'\times f(r')\mathbf{X}_{l'm'}^*(\Omega)\cdot
(\mathbf{1}-\mathbf{\hat r\hat r})\cdot
\bm{\nabla}\times g(r)\mathbf{X}_{lm}(\Omega)&=&\delta_{ll'}\delta_{mm'}
\frac1{rr'}\frac\partial{\partial r'}\left(r'f(r')\right)\frac\partial{\partial
r}\left(r g(r)\right).
\eea
\end{subequations}
The trace in Eq.~(\ref{fe})
is carried out by expanding the logarithm, doing the trace in each 
order, and resumming to get
\be
F=\frac{T}2\sum_{n=-\infty}^\infty\sum_{l=1}^\infty (2l+1)
\left[\ln(1+\lambda\zeta_n^2a^2 g^0_l(a,a))
+\ln\left(1-\lambda\partial_r r\partial_{r'} r'g^0_l(r,r')\right)\big|_{r=r'=a}
\right].
\ee
The free reduced spherical Green's function is
\be
g^0_l(r,r')=\frac1{\kappa rr'}s_l(|\zeta_n| r_<)e_l(|\zeta_n| r_>),
\ee
where $r_{<(>)}$ represents the lesser or greater of $r$
and $r'$, and the modified Ricatti-Bessel functions are
\be
s_l(x)=\sqrt{\frac{\pi x}2}I_{l+1/2}(x),\qquad e_l(x)=\sqrt{\frac{2x}\pi}
K_{l+1/2}(x),
\ee
which have a Wronskian equal to one.

The final form of the expression we must evaluate for the free energy of
a semitransparent sphere is defined by (1) inserting point-splitting in 
(imaginary) time, with parameter $\tau$,
 and in the spatial directions transverse to the normal
of the sphere, with angle parameter $\delta$, 
 and (2) by using the ``plasma-model'' dispersion relation
for the coupling, $\lambda=\lambda_0/\zeta_n^2 a$, where $\lambda_0$ is a
dimensionless constant.  
These two processes are precisely those followed in Ref.~\cite{ly}.
The regulated free energy is ($x=|\zeta_n| a$) 
\be
F=\frac{T}2\sum_{n=-\infty}^\infty e^{i\zeta_n\tau}\sum_{l=1}^\infty (2l+1)
P_l(\cos\delta)\left[\ln\left(1+\lambda_0\frac{e_l(x)s_l(x)}x\right)+
\ln\left(1-\lambda_0\frac{e'_l(x)s'_l(x)}x\right)\right].\label{gfe}
\ee
The first term here is the transverse electric (TE) free energy and the second is
the transverse magnetic (TM).
These two contributions, which reduce to the familiar result for a
perfectly conducting shell in the limit $\lambda_0\to \infty$, are just those
discussed earlier in Ref.~\cite{casimir-physics} [see Eq.~(3.96) there], 
with the following
identifications of the coupling constants there: $\lambda^{\rm TE}
=\lambda_0 a$,
$\lambda^{\rm TM}=-\lambda_0 a/x^2$. In the notation of 
Ref.~\cite{Parashar:2012it}, $\lambda_0=\zeta_p a$.  It is important to keep
this $a$ dependence in computing the self-stress on the sphere from 
$\mathcal{S}=-\frac{\partial}{\partial a}F$; but here we are interested in the
entropy,
$S=-\frac\partial{\partial T}F$, 
 so for notational convenience we keep the coupling as $\lambda_0$.

\section{Strong Coupling}
\label{sec4}
Since the spherical-shell entropy
problem has mostly been considered in the perfectly-conducting
limit, we begin with that situation.
In strong coupling, $\lambda_0\to \infty$, the free energy reduces to
\be
F_\infty=\frac{T}2\sum_{n=-\infty}^\infty e^{i\zeta_n\tau}\sum_{l=1}^\infty
(2l+1)P_l(\cos\delta)\ln\frac{e_l(x)s_l(x)e'_l(x)s'_l(x)}{x^2},\label{finf}
\ee
where the coupling has disappeared because
\be
\sum_{l=1}^\infty (2l+1)P_l(\cos\delta)=-1 \quad (\delta\ne0)\quad \mbox{and}
\quad\sum_{n=-\infty}
^\infty e^{inx}=0 \quad (x\ne 0 \mod 2\pi).\label{completeness}
\ee
To isolate divergences we use the uniform asymptotic expansion (UAE) 
\cite{nist},
which gives for the leading behavior of the logarithm \footnote{
This is an astoundingly good approximation.  Even for $l=1$ the discrepancy
between the two sides of Eq.~(\ref{uaesc}) is within 0.1\% for all $x$.}
\be
\ln e_l(x)s_l(x)e'_l(x)s'_l(x)\sim -\ln4-\frac{t^6}{4\nu^2}+
\frac{t^6}{32\nu^4}(4-54t^2+120 t^4-71 t^6)+O(\nu^{-6}),\quad \nu\to\infty,
\label{uaesc}
\ee
where $x=\nu z$, $t=(1+z^2)^{-1/2}$, and $\nu=l+1/2$.  Note that there
are no odd orders of $\nu^{-1}$ in this expansion.
This expression is
actually valid at $n=0$ where $t=1$ [see Eq.~(\ref{n0sc}) below].

\subsection{Leading term}
The leading term in the expansion of Eq.~(\ref{finf}),
\be
F^{(0)}_\infty
=\frac{T}2\sum_{n=-\infty}^\infty e^{i\zeta_n\tau}\sum_{l=1}^\infty (2l+1)
P_l(\cos\delta)(-\ln4x^2),
\ee
would be thought as a priori irrelevant, since it doesn't refer to the sphere, 
in view of Eq.~(\ref{completeness}).
However, if we take it seriously, we write it as
\be
F^{(0)}_\infty=
\frac{T}2\sum_{n=-\infty}^\infty e^{i\zeta_n\tau}\ln 4\zeta_n^2a^2.
\label{f0}
\ee
The $n=0$ term is not defined here, so we regularize this infrared divergence
by replacing $\zeta_n^2$ by $\zeta_n^2+\mu^2$, where dimensionlessly, 
 $p=\mu a$, $\mu$ being a
photon ``mass.''\footnote{In Ref.~\cite{Bordag:1998vs}, the case of a massive field is 
considered.  As noted there, when the $a_2$ heat-kernel coefficient is nonzero,
as is the case here, the massless limit cannot be taken.}
We will assume $p$ is smaller than any other scale 
in the problem. Thus, the $n=0$ term is simply
\be
F_{\infty,n=0}^{(0)}=\frac T2\ln 4p^2,
\ee
but for $n\ne0$, $p$ will be neglected.  Then, the sum can be expressed
in terms of polylogarithms, where we now abbreviate $\alpha=2\pi a T$:
($\tilde\tau=\tau/a$)
\be
F^{(0)}_{\infty,n\ne0}
=2T\frac\partial{\partial\beta}\sum_{n=1}^\infty(2\alpha n)^\beta\cos
\alpha n\tilde\tau\bigg|_{\beta=0}=2T\Re \frac{\partial}{\partial \beta}\left[
(2\alpha)^\beta \Li_{-\beta}\left(e^{i\alpha\tilde\tau}\right)
\right]\bigg|_{\beta\to0}.
\ee
At $\beta=0$, $\Li_0(z)=z/(1-z)$, while it can be shown for small $\tau$ that
\be
\frac\partial{\partial\beta}\Li_{-\beta}\left(e^{i\alpha\tilde\tau}\right)
\bigg|_{\beta=0}
=\frac1{i\alpha\tilde\tau}[\ln(-i\alpha\tilde\tau)+\gamma]-\frac12\ln 2\pi, 
\ee
so we find that
\be
F_\infty^{(0)}=\frac T2\ln 4p^2-\frac1{2\tau}-T\ln\frac{2\alpha}{2\pi}=
-\frac1{2\tau}+T\ln\frac{2\pi p}\alpha
=-\frac1{2\tau}+T\ln\frac{\mu}{T}.\label{n=0sc}
\ee

The same result is easily obtained by use of the Euler-Maclaurin sum formula,
\be
\sum_{n=0}^\infty{}' g(n)=\int_0^\infty dn\, g(n)-\sum_{k=1}^\infty 
\frac{B_{2k}}{(2k)!}g^{(2k-1)}(0),\label{em}
\ee
where the prime means that the $n=0$ term is counted with half weight.  This
formula
 provides a formal asymptotic expansion in terms of the Bernoulli numbers.
Even more simply, a corresponding answer is found when we use analytic
regulation (we will return to this technique later in Sec.~\ref{sec:hit}), 
defining
\bea
F^{(0)}_{\infty,n\ne0}&=&-T\sum_{n=1}^\infty (2l+1)\ln 4n^2\alpha^2
=\frac{11}6 T\frac\partial{\partial \beta}\sum_{n=1}^\infty (2n\alpha)^\beta
\bigg|_{\beta\to0}\nn\\
&=&\frac{11}6T[\zeta(0)\ln 2\alpha-\zeta'(0)]=-\frac{11}{12}T\ln
\frac{2\alpha}{2\pi},
\eea
or
\be
F^{(0)}_\infty=\frac{11}{12}T\ln\frac{2\pi p}\alpha.\label{irfe}
\ee
which uses the replacement $\sum_{l=1}^\infty (2l+1)\to -\frac{11}{12}$ 
rather than $-1$ obtained from point splitting.  Aside from this fact
the ultraviolet finite answer obtained is the same as that given in Eq.~(\ref{n=0sc}).
Evidently, with the infrared sensitive term included, this contribution is
independent of the sphere's radius (obvious a priori); that fact, together
with the sensitivity of the coefficient to the choice of regulation scheme,
seems to argue that this contribution is unphysical, and should be disregarded
(subtracted).

\subsection{Low-temperature behavior}

To see the low-temperature dependence of the free energy, 
and thus the entropy, return to the exact
strong coupling formula (\ref{finf}).  As we see from the Euler-Maclaurin
formula, to get a temperature correction to the zero-temperature energy
we need a  contribution
odd in $n$.  In fact, the lowest-order odd term in the logarithm 
occurs for
$l=1$, where $\ln(-e_1e'_1s_1s'_1)\sim \dots -\frac13 x^3+\dots$, where the 
leading omitted terms
 are even in $x$.  Then the Euler-Maclaurin formula immediately leads to
 the leading
low-temperature correction to the free energy
\be
\Delta F_\infty=T\frac{B_4}{4!}3\cdot2 (2\pi T a)^3=-\frac{(\pi a)^3}{15}T^4,
\quad aT\ll1,
\label{lowtsc}
\ee
as first found in Ref.~\cite{bd}.  [See also Ref.~\cite[Sec.~9.5]{bkmm}.]

\subsection{High-temperature behavior}
To get the high-temperature behavior, we need to consider $n=0$ separately from
 $n\ne0$. The former is
\be
F_{\infty,n=0}
=\frac{T}2\sum_{l=1}^\infty (2l+1)P_l(\cos\delta)\ln\left(1
-\frac{1}{(2l+1)^2}\right)
=-\frac{T}2\sum_{k=1}^\infty \frac1k\sum_{l=1}^\infty \frac{P_l(\cos\delta)}
{(2l+1)^{2k-1}}.
\label{n0sc}
\ee
The $k=1$ term here is divergent as $\delta\to0$. 
The balance of $F_{\infty,n=0}$ sums to
\be
F_{\infty,n=0}\to-T\left(\frac12-\frac14\gamma+\frac7{12}\ln 2-3\ln G\right)
=-\frac{T}2(0.027537),\label{n0scs}
\ee
where $G$ is Glaisher's constant, $\ln G=\frac1{12}-\zeta'(-1)$.
For $n\ne0$, the 
$O(\nu^{-2})$ term in the UAE expansion of
the logarithm seen in Eq.~(\ref{uaesc}) is given as 
\be
F_\infty^{(2)}=-T\sum_{n=1}^\infty\cos\zeta_n\tau\sum_{l=1}^\infty
\frac{P_l(\cos\delta)}{2l+1}
\frac{\nu^6}{(\nu^2+\zeta_n^2a^2)^3},
\ee
where the sum on $n$ is readily carried out to yield (for simplicity, we
have set $\tau=0$)
\be
F_\infty^{(2)}=-\frac{T}2\sum_{l=1}^\infty P_l(\cos\delta)\left\{-\frac1{2\nu}+
\frac1{32 aT}
\left[3\coth\frac\nu{2aT}+\frac32\frac\nu{aT}\csch^2\frac\nu{2aT}+\frac12
\left(\frac\nu{aT}\right)^2\coth\frac\nu{2aT}\csch^2\frac\nu{2aT}\right]
\right\}.\label{nsc}
\ee
The first term in braces in this expression precisely cancels the $k=1$ term in
 Eq.~(\ref{n0sc}).
For large $\nu/2aT$ (this is the low-temperature limit)
 the hyperbolic cotangent in Eq.~(\ref{nsc}) tends to one, 
so we should remove that limiting term, which amounts to the zero-energy
term,
\be
F_\infty^{T=0}
=-\frac{T}2\sum_{l=1}^\infty P_l(\cos\delta)\frac3{32aT}=\frac3{64a}
-\frac3{64 a\delta}.\label{t=0}
\ee
In the same limit, the balance of Eq.~(\ref{nsc}) vanishes exponentially fast.
The finite part of this is, as noted in Ref.~\cite{mds}, within 2\% of the
 exact
Boyer result \cite{boyer,bd,mds} for a perfectly conducting spherical shell,
$E_B=-0.04617/a$.  The divergent term appearing here does not appear in other
ways of regulating the zero-point energy.  Including higher terms in the UAE,
and computing the remainder, indeed gives exactly the Boyer result, at zero
 temperature.

For high temperature,
we evaluate what is left in Eq.~(\ref{nsc})  via the Euler-Maclaurin formula.
The integral gives the leading term:
\begin{subequations}
\bea
F_\infty^{\prime\prime,T\to\infty}
&=&-\frac{T}{32}\int_{3/4aT}^\infty dx[3(\coth x-1)+3x \csch^2x
+2 x^2\coth x\csch^2x]\nn\\
&=&-\frac{T}{32}\left[\frac3{4aT}\left(3+5 \coth\frac{3}{4aT}
+\frac3{4aT}\csch^2\frac3{4aT}\right)
-8\ln\left(2\sinh\frac3{4aT}\right)\right]\\
&\to&-\frac{T}{16}\left(3-4\ln\frac3{2aT}\right)-\frac9{128a} +O(T^{-5}).
\label{sclog}
\eea
\end{subequations}
The last limit assumes $aT\gg 1$. 
 The remaining terms in the Euler-Maclaurin 
series are easily evaluated using Borel summation:
\be
\sum_{n=0}^\infty \frac{B_{n+1}}{n+1} x^n=\int_0^\infty dt e^{-t}\frac1{xt}
\left(\frac{xt}{e^{xt}-1}-1\right)
=-1-\frac1x \ln x-\frac1x \psi\left(\frac1x\right),
\ee
where $\psi$ is the digamma function.
This yields the final contribution to the free energy:
\be
F^{\prime\prime\prime,T\to\infty}_\infty
=\frac{T}4\left[\ln\frac23+\psi\left(\frac32\right)\right]+\frac3{128a}.
\ee
Adding this result to that in Eqs.~(\ref{n0scs}), (\ref{t=0}), 
and (\ref{sclog}), we see
that the finite temperature-independent part cancels, leaving for  
the high-temperature limit apart from a divergent constant
\be
F_{\infty}^{T\to\infty}\sim T\left(-\frac14\ln aT-\frac3{16}-\frac{13}{12}
\ln 2+3\ln G\right)=-\frac{T}4(\ln aT+0.768584),\quad aT\gg1,\label{hitpc}
\ee
a result first obtained by Balian and Duplantier \cite{bd}. (See also 
 Teo \cite{teo} and Bordag et al.~\cite{bkmm}.) [To understand this high-temperature limit better, 
 we will break this up into TE and TM parts in Sec.~\ref{sec:sctetm}.]

To obtain the next term in the high-temperature expansion, we consider the
order $\nu^{-4}$ term in Eq.~(\ref{uaesc}).  This term can be written as
\be
F_\infty^{(4)}=-\frac{T}{16}\sum_{l=1}^\infty \nu^3 g(y),
\ee
where, with $y=\nu/2Ta$ and $\alpha=2\pi Ta$,
\be
g(y)=\frac1{2(\alpha/\pi)^6}
\left[2\left(\frac{d}{dy^2}\right)^2+9y^2\left(\frac{d}{dy^2}
\right)^3+5y^4\left(\frac{d}{d y^2}\right)^4+\frac{71}{120}y^6
\left(\frac{d}{d y^2}\right)^5\right]\left[-\frac1{y^2}+\frac1y\coth y\right].
\ee
The leading contribution comes from the first, integral, term in the Euler-%
Maclaurin formula:
\be
F_\infty^{(4),T\to\infty}
\sim -\frac{T}{16}\left(\frac\alpha\pi\right)^4
\int_{3\pi/2\alpha\to 0}^\infty dx\, x^3 g(x)
=-\frac1{3840 a^2 T},
\ee
or altogether,
\be
F^{T\to\infty}_\infty\sim -\frac{T}4(\ln aT+0.7686)
-\frac1{3840a^2 T},\label{htscbd}
\ee 
again as first derived in Ref.~\cite{bd}.

\section{Finite Coupling Behaviors}
\label{sec5}
Now let's return to Eq.~(\ref{gfe}) with finite coupling $\lambda_0$, and
approximate the logarithm using the UAE:
\be
\ln\left[\left(1+\frac{\lambda_0}x e_l(x)s_l(x)\right)
\left(1-\frac{\lambda_0}x e_l'(x)s_l'(x)
\right)\right]\sim \sum_{k=1}^\infty \frac{a^{(k)}(t)}{(2\nu)^k},\quad \nu\gg1,
\label{uaelog}
\ee
where the first four expansion coefficients are
\begin{subequations}
\bea
a^{(1)}(t)&=&2\lambda_0 t,\\
a^{(2)}(t)&=&-\lambda_0^2 t^2,\\
a^{(3)}(t)&=&\frac{\lambda_0}3(-3t^7+2\lambda_0^2 t^3),\\
a^{(4)}(t)&=&\frac{\lambda_0^2}2(2t^8-\lambda_0^2 t^4),
\eea
\end{subequations}
with again $t=(1+z^2)^{-1/2}$,  $x=\nu z$.  Here we have
dealt with the infrared divergence in the TM contribution by replacing
$1/z^2 t\to t$, as discussed in Ref.~\cite{prachi}, because this substitution
does not change the ultraviolet behavior. The idea was that the error
introduced is compensated by the remainder, and this substitution is sufficient
to capture the divergence structure.
 However, this would not be expected
to be valid for the $n=0$ term, so we will return to this point later, when
discussing the temperature dependence.
Note that this expansion is not a power-series expansion in the coupling 
$\lambda_0$.

The first-order term in this expansion is
\be
F^{(1)}=2\lambda_0T\sum_{n=0}^\infty {}'\cos\zeta_n\tau\sum_{l=1}^\infty P_l
(\cos\delta)\frac1{\sqrt{1+\zeta_n^2a^2/\nu^2}}.
\ee
For low temperature, we again evaluate the sum on $n$ 
using the Euler-Maclaurin sum formula (\ref{em}).
The integral term there is all there is, because the $n$ summand
is even about $n=0$. (This will result in no temperature dependence being
revealed, as we saw in the previous section.)
 That integral is immediately seen to be
\be
F^{(1)}=\frac{\lambda_0}{2\pi a}\sum_{l=1}^\infty (2l+1)P_l(\cos\delta)
K_0(\nu\tilde\tau),\quad \tilde\tau=\tau/a.
\ee
Since only the $\tau$ cutoff is essential here, we set $\delta\to0$ and have
\be
F^{(1)}=\frac{\lambda_0}{\pi a}\sum_{l=1}^\infty \nu K_0(\nu\tilde\tau).
\ee
This sum, in turn, may be evaluated using Euler-Maclaurin around $l=1$.
The integral term gives
\be
F^{(1a)}=\frac{\lambda_0}{\pi a}\left[\frac1{\tilde\tau^2}+\frac9{16}
\left(-1+2\gamma+2\ln\frac{3\tilde\tau}4\right)\right].
\ee
The remainder terms involve, with $g(l)=\nu K_0(\nu\tilde\tau)$,
\begin{subequations}
\bea
\frac12 g(1)&=&-\frac34\left(\ln\frac{3\tilde\tau}4+\gamma\right),\\
g'(1)&=&-\gamma-1-\ln\frac{3\tilde\tau}4,\\
g^{(2k-1)}(1)&=&\frac{(2k-3)!}{(3/2)^{2k-2}},\quad k>1.
\eea
\end{subequations}
The resulting series is Borel-summable:
\bea
F(x)&=&x\sum_{k=2}^\infty B_{2k}\frac{(2k-3)!}{(2k)!}x^{2k-3}=
x\int_0^\infty dt \frac{e^{-t}}{(xt)^3}\left[\frac{xt}{e^{xt}-1}-B_0-B_1 xt
-\frac{B_2}2(xt)^2\right]\nn\\
&=&\frac1{12x^2}\left[-9-6x+(6+6x+x^2)(\gamma+\ln x)-12x^2(1-\gamma)
\zeta\left(-1,1+\frac1{x}\right)-12 x^2\zeta^{(1,0)}\left(-1,1+\frac1x\right)
\right],
\eea
where the integral is 
evaluated at $x=2/3$,
 by analytically continuing in the power of $xt$ in the denominator.
The numerical value $F(2/3)=-0.00058434$.  
Adding together the components, we obtain
\be
F^{(1)}=\frac{\lambda_0}{\pi a}\left[\frac1{\tilde\tau^2}+\frac{11}{24}\ln
\tilde\tau-0.345879\right].
\ee

The second term in the UAE gives
\be
F^{(2)}=-\lambda_0^2 T\sum_{l=1}^\infty \frac{P_l(\cos\delta)}{2\nu}
\sum_{n=0}^\infty{}'\cos\zeta_n\tau\frac1{1+\zeta_n^2a^2/\nu^2}.
\ee
Evaluating the $n$ sum again by the Euler-Maclaurin formula, the integral
there gives
\be
F^{(2)}=-\frac{\lambda_0^2}{8a}\sum_{l=1}^\infty P_l(\cos\delta)e^{-\nu\tilde
\tau}=\frac{\lambda_0^2}{8a}\left(1-\frac1\Delta\right),\quad \Delta=
\sqrt{\delta^2+\tilde\tau^2}.
\ee
There are no remainder terms, because the $n$-summand is even.

The third term
\be
F^{(3)}=-\frac{\lambda_0T}3\sum_{n=0}^\infty{}'\cos\zeta_n\tau\sum_{l=1}^\infty
\frac{P_l(\cos\delta)}{(2\nu)^2}\left[-\frac3{(1+\zeta_n^2a^2/\nu^2)^{7/2}}
+\frac{2\lambda_0^2}{(1+\zeta_n^2a^2/\nu^2)^{3/2}}\right],
\ee
is most easily evaluated when $\tau=0$.  As before the $n$-sum can be replaced
by an integral, and the sum on $l$ is easily carried out, with the result
\be
F^{(3)}=\frac{\lambda_0}{30\pi a}(4-5\lambda_0^2)\left(1-\frac32\ln 2
+\frac12\ln\delta\right).
\ee  
The fourth term, and those thereafter, are finite:
\be
F^{(4)}=\frac{\lambda_0^2}{128 a}\left(\frac{\pi^2}8-1\right)(5-4\lambda_0^2).
\ee
The above analysis, based on the uniform asymptotic expansion, 
and the Euler-Maclaurin summation formula, exhibits no
temperature dependence in the free energy.  This is because the $n$-summand
is even in $n$, so the Euler-Maclaurin formula (or the Abel-Plana formula)
allows the sum to be replaced
by an integral, which precisely corresponds to the zero-temperature energy.

\section{Lowest-order coupling contribution to the free energy}
\label{sec:exactlow}

As in strong coupling, to get the low-temperature correction,
we must return to the exact expression (\ref{gfe}).  In this
section we will consider low orders in the coupling, which can
be treated exactly.
We will first consider the first-order in $\lambda_0$  contributions.

\subsection{TE $O(\lambda_0)$ behavior}
\label{sec:telambda}
The TE contribution to the free energy in Eq.~(\ref{gfe}) may be expanded 
to first order in $\lambda_0$ as
\be
F^{\rm{TE}}_{(1)}=\lambda_0\frac{T}2\sum_{n=-\infty}^\infty e^{i\zeta_n\tau}
\sum_{l=1}^\infty (2l+1)P_l(\cos\delta)\frac{e_l(x)s_l(x)}x.
\ee
We can evaluate this exactly.  We use the summation theorem \cite{Klich:1999df}
\be
\sum_{l=0}^\infty (2l+1)P_l(\cos\delta)e_l(x)s_l(y)=\frac{xy}\rho e^{-\rho},
\quad \rho=\sqrt{x^2+y^2-2xy\cos\delta}\label{klich-formula}
\ee
to evaluate the $l$-sum as
\be
\sum_{l=1}^\infty (2l+1)P_l(\cos\delta)e_l(x)s_l(x)=\frac{x}u 
e^{-|x|u}-e_0(x)s_0(x),
\ee
where $u=\sqrt{2(1-\cos\delta)}\approx\delta$.  
Then the sum over Matsubara frequencies is ($\alpha=2\pi a T$, 
$\tilde\tau=\tau/a$)
\be
F^{\rm{TE}}_{(1)}=\lambda_0\frac{T}2\sum_{n=-\infty}^\infty 
e^{in\alpha\tilde\tau}
\left[\frac1u e^{-|n|\alpha u}+\frac1{2|n|\alpha}
\left(e^{-2|n|\alpha}-1\right)\right].
\ee
This is readily evaluated to be
\be
F_{(1)}^{\rm{TE}}=\frac{\lambda_0}{2\pi a}\left(\frac1{u^2+\tilde\tau^2}
+\frac12\ln\frac{\tilde\tau}2
+\frac1{12}\alpha^2-\frac12\ln\frac{\sinh\alpha}\alpha\right).\label{te1exact}
\ee
The free energy diverges as $\tau$ and $\delta$ tend to zero, but the entropy 
is finite:
\be
S_{(1)}^{\rm TE}=-\frac{\partial F^{\rm TE}}{\partial T}
=-\lambda_0\left(\frac\alpha6+\frac1{2\alpha}-\frac12
\coth\alpha\right).\label{ste1}
\ee
For low temperature, the entropy tends to $-\lambda_0 \alpha^3/90$, 
a result which will be rederived below, while for high temperature,
\be
S_{(1)}^{\rm TE}\sim -\lambda_0\left(\frac\alpha6-\frac12+\frac1{2\alpha}\right).
\ee

\subsection{TE $O(\lambda^2_0)$ behavior}
We can also exactly calculate the TE  $O(\lambda_0^2)$ behavior.  Squaring the
identity (\ref{klich-formula}) and integrating over angles, we obtain
($w=2x\sqrt{2(1-\cos\theta)}$, where $\theta$ is the angle in the sum rule) 
\be
\sum_{l=0}^\infty (2l+1)e_l^2(x)s_l^2(x)=\frac{x^2}2\int_0^{4x}\frac{dw}w
e^{-w},
\ee  which, unfortunately, loses the angular point-splitting regulator.  So now
the second-order TE free energy is given by the expression
\be
F^{\rm TE}_{(2)}=-\frac{\lambda_0^2 T}4\sum_{n=0}^\infty{}'\cos n\alpha\tilde
\tau\left[\int_0^{4n\alpha}\frac{dw}w e^{-w}-\frac{2}{x^2}
e_0^2(x)s_0^2(x)\right].\label{l2te}
\ee
Because this expression exhibits no ultraviolet divergence, we can set
$\tau$ to zero.
Let us consider the second term (the $l=0$ contribution), first. 
Since $e_0^2(x)s_0^2(x)=e^{-2x}\sinh^2(x)$, it is immediately evaluated as
\be
F^{\rm TE}_{a}=\frac{\lambda_0^2\alpha}{8\pi a}\left\{1
+\frac1{2\alpha^2}\left[
\frac{\pi^2}6-2\mbox{Li}_2\left(e^{-2\alpha}\right)+\mbox{Li}_2
\left(e^{-4\alpha}\right)\right]\right\}.\label{fte2l0}
\ee

Let us similarly consider the low temperature limit of the first term in
Eq.~(\ref{l2te}), so we use the Euler-Maclaurin formula to evaluate it,
but because the summand is singular at $n=0$, we do so about $n=1$:
\be
F^{\rm TE}_{b}=-\frac{\lambda_0^2 \alpha}{8\pi a}\left[\int_0^\infty dn \, f(n)
-\int_0^1 dn\,f(n)+\frac12f(0)+\frac12f(1)-\sum_{k=1}^\infty 
\frac{B_{2k}}{(2k)!} f^{(2k-1)}(1)\right].\label{emn1}
\ee
We can disregard the first term here, the integral from 0 to $\infty$, because
when the variable is changed from $n$ to $n\alpha$, the integral is seen to be 
independent of $T$, and hence does not contribute to the entropy.  The second
integral from 0 to 1 is 
\be
\frac{\lambda_0^2\alpha}{8\pi a}\int_0^1 dn\,f(n)=
\frac{\lambda_0^2 \alpha}{8\pi a}\left[\int_\eta^{4\alpha}\frac{dw}w e^{-w}+
\frac1{4\alpha}\left(e^{-4\alpha}-1\right)\right]
=\frac{\lambda_0^2\alpha}{8\pi a}\left[\frac1{4\alpha}\left(e^{-4\alpha}-
1\right)+\Ei(-4\alpha)-\Ei(-\eta)\right],
\ee where $\Ei$ is the exponential integral function, and
we have regulated the divergence at $\cos\theta=1$ ($w=0$) by inserting
a small positive number $\eta$.
To evaluate the $f(0)$ term we again need to insert the photon mass parameter $p$,
and then we obtain
\be
\frac{\lambda_0^2\alpha}{8\pi a}\frac12f(0)=\frac{\lambda_0^2\alpha}{16\pi a}
\ln\frac{4p}\eta.
\ee
The $\frac12 f(1)$
term is
\be
-\frac{\lambda_0^2\alpha}{16\pi a}\int_\eta^{4\alpha}\frac{dw}w e^{-w}=
-\frac{\lambda_0^2 \alpha}{16\pi a}\left[
\Ei(-4\alpha)-\Ei(-\eta)\right].
\ee
The terms in the Bernoulli series are also readily worked out to all orders:
\be
f^{(2k-1)}(1)=\Gamma(2k-1,4\alpha).
\ee
The leading terms in the Bernoulli expansion are evaluated
by Borel summation,
\be
-\sum_{k=1}^\infty \frac{B_{2k}}{(2k)!}(2k-2)!=-\int_0^\infty \frac{dt}{t^2}
e^{-t}\left(\frac{t}{e^t-1}-1+\frac{t}2\right)=\frac12\ln2\pi-1.
\label{kay}
\ee
Then, adding all these components, we find  through $O(\alpha^4)$ the 
second-order TE free energy to be
\be
\Delta F^{\rm TE}_{(2)}=
-\frac{\lambda_0^2}{8\pi a}\left\{\frac\alpha2\ln\frac{2\pi p}{\alpha}
+\frac{\alpha^4}{270}\right\},\label{dfl2}
\ee
where the $\Delta$ symbol signifies that temperature-independent constants
have been dropped.
Although the $\alpha^2$ and $\alpha^3$ terms canceled, as has the colinear cutoff
$\eta$, there persists a linear term in $T$ 
(dependent on the photon-mass infrared cutoff) and a $T\ln T$ term,
both of which, if present, would violate the Nernst
heat theorem. We have  seen precisely such terms appearing in the 
strong-coupling limit, Eq.~(\ref{n=0sc}), and will see additional power
divergences  in the TM contribution to 
the free energy already in order $\lambda_0$.  As we will argue in the next
section and subsequently, such divergences are 
probably to be omitted.  In particular, we will not see this term when
we consider the general low-temperature expansion in Sec.~\ref{sec:telow}.

\begin{figure}
\includegraphics{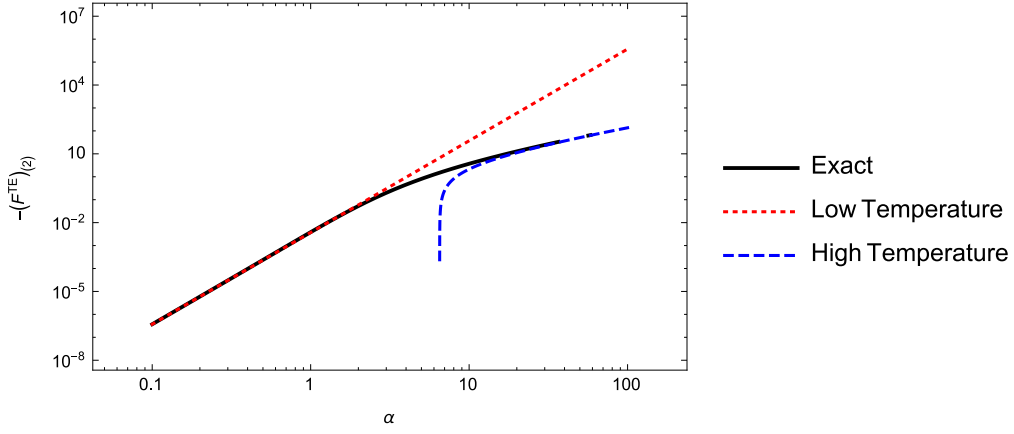}
\caption{\label{figt2order} The negative of the TE free energy in second order 
in $\lambda_0$, apart from a factor of
$\lambda_0^2/(8\pi a)$, plotted as a function of $\alpha=2\pi a T$.
The solid line shows the numerically computed
free energy (\ref{fe2order}) compared with the
low-temperature approximation (dotted curve) and the high-temperature
approximation (dashed curve), given in Eq.~(\ref{2llimits}).}
\end{figure}

We can easily extend this calculation to all orders in $\alpha$.  The key
observation is that in $O(\alpha^{2k})$, $k\ge1$ , 
all contributions cancel except that
from the $B_{2k}$ term in the Bernoulli sum and the contribution from 
the $l=0$ term (\ref{fte2l0}). 
This is a consequence of the identity
\be
\sum_{k=0}^{n-1}\left(\begin{array}{c}
n\\k\end{array}\right)B_k=0,\quad n>1.
\ee
Adding up the remainder, leads to the 
temperature-dependence of the free energy in second order:
\be
\Delta F^{\rm TE}_{(2)}=\frac{\lambda_0^2}{8\pi a}\left\{\alpha+\frac1{2\alpha}
\left[\frac{\pi^2}6-2\mbox{Li}_2(e^{-2\alpha})+\mbox{Li}_2(e^{-4\alpha})\right]
-\frac14 h(4\alpha)\right\},\label{fe2order}
\ee
where
\be
h(x)=x\int_0^x
\frac{dt}{t^2}\left[\frac{t}{e^t-1}-1+\frac12 t\right],
\label{effofalpha}
\ee
which has the limits
\be
h(x)\sim\frac{x^2}{12} -\frac{x^4}{2160},\quad x\ll1,\quad
h(x)\sim \frac{x}2\ln x-Ax, \quad A=0.63033,\quad x\gg 1.
\ee
The limiting behaviors of the free energy are
\be \Delta F^{\rm TE}_{(2)}\sim\frac{\lambda_0^2}{8\pi a}\left\{
\begin{array}{cc}
-\frac{\alpha^4}{270},&\alpha\ll1,\\
1.63033 \alpha-\frac12\alpha\ln 4\alpha,& \alpha\gg1.
\end{array}\right.\label{2llimits}
\ee
The second-order TE free energy
is shown in Fig.~\ref{figt2order}.  Note that because $\Delta F^{\rm TE}_{(2)}$
has negative slope, the corresponding entropy contribution is positive,
unlike the first-order contribution.

\subsection{TM $O(\lambda_0)$ behavior}  
\label{sec:tmweak}
Extracting the weak-coupling behavior of the TM contribution is considerably
more subtle.  The expression we need to evaluate is
\be
F^{\rm TM}_{(1)}=-\lambda_0T\sum_{n=0}^\infty{}'\cos n\alpha\tilde\tau 
\sum_{l=1}^\infty
(2l+1)P_l(\cos\delta)\frac1{n\alpha} e_l'(n\alpha) s_l'(n\alpha).
\ee
We first note that the $n$ sum is divergent because of an infrared divergence
at $n=0$. To regulate this, we again insert the small photon mass parameter
$p$.  Then the $n=0$ contribution is immediately worked
out:
\be
F^{\rm TM}_{n=0}=\frac{\lambda_0\alpha}{4\pi a }\left(-\frac{1+u^2/4}{p^2u^3}
+1+\frac{u^2+\tilde\tau^2}{2u^3}\right).\label{ftm0}
\ee
For the rest, we ignore the photon mass, and use
the summation formula (\ref{klich-formula}) to write $F^{\rm TM}_{n\ne0}
=F^{\rm TM}_{a}+F^{\rm TM}_{b}$, where
\begin{subequations}\label{Fab}
\bea
F^{\rm TM}_{a}
&=&\lambda_0 T\sum_{n=1}^\infty \cos n \alpha \tilde\tau \frac{
e_0'(n\alpha)s_0'(n \alpha)}{n\alpha},\\
F^{\rm TM}_{b}&=&-\lambda_0 T\sum_{n=1}^\infty \cos n\alpha\tilde \tau
\frac{e^{-u n \alpha}}{4 u^3(n\alpha)^2}[4+u^2+u(4-3u^2)n\alpha
+u^4(n\alpha)^2].
\eea
\end{subequations}
Because $e_0'(x)s_0'(x)=-\frac12\left(1-e^{-2x}\right)$, we can readily
evaluate the subtracted $l=0$ term:
\be
F^{\rm TM}_a=\frac{\lambda_0}{4\pi a}[\ln\tilde\tau+\ln\alpha-\alpha
+\ln(2\sinh\alpha)].\label{ftma}
\ee
The $n$ sums in the remaining part $F^{\rm TM}_{b,n\ne0}$ are straightforward, 
leading to, for small $u$ and $\tilde\tau$,
\bea
F^{\rm TM}_b&=&-\frac{\lambda_0}{8\pi a}\bigg\{\frac23\frac{\pi^2}{u^3\alpha}
-\frac4{u^2}\left(1+\frac{\tilde\tau}u \arctan\frac{\tilde \tau}u\right)
+\frac1u\left[\left(\frac{\tilde\tau^2}{u^2}+1\right)\alpha
+\frac{\pi^2}{6\alpha}\right]\nn\\
&&\quad\mbox{}+2\ln(u^2+\tilde\tau^2)\alpha^2
-\frac{\tilde\tau}u\arctan\frac{\tilde
\tau}{u}-1+\frac1{1+\tilde\tau^2/u^2}-\frac{\alpha^2}9\bigg\}.\label{ftmb}
\eea
When Eqs.~(\ref{ftm0}), (\ref{ftma}), and (\ref{ftmb}) are combined, we are
left with divergent terms that depend on temperature, as well as a finite
remainder:
\bea
F^{\rm TM}_{(1)}&=&-\frac{\lambda_0}{2\pi a}
\bigg\{\frac1{u^3}\left(\frac{\pi^2}{6\alpha}+\frac{\alpha}{2p^2}\right)
\left(1+\frac{u^2}4\right)
+\bigg[-\frac1{u^2}\left(1+\frac{u^2}4\right)\left(1+\frac{\tilde
\tau}u\arctan\frac{\tilde\tau}u\right)\nn \\
&&\qquad\mbox{}-\frac12\ln\tilde\tau+\frac12\ln(u^2
+\tilde \tau^2)+\frac14\frac1{1+\tilde\tau^2/u^2}\bigg]
-\frac12\ln\frac{2\sinh\alpha}{\alpha}-\frac{\alpha^2}{36}
\bigg\}.\label{totalF1}
\eea
Interestingly, with this way of regulating the infrared divergence, the 
$\tau$-dependent  and finite terms linear in $\alpha$ have canceled between
$F^{\rm TM}_{n=0}$ and $F^{\rm TM}_{n\ne0}$. 

\begin{figure}
\centering
\includegraphics{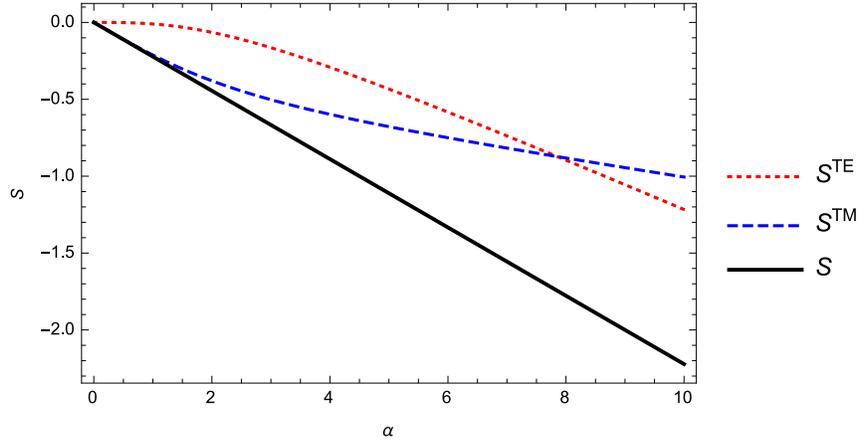}
\caption{\label{figSx} 
The TE and ``renormalized'' 
 TM contributions to the entropy in first
order in the coupling, plotted as a function of $\alpha=2\pi a T$.
  Both the TE and the TM  contributions are negative,
leading to a sum which is negative  and linear in the 
temperature.}
\end{figure}

The entropy is obtained from the free energy by differentiating with respect
to $\alpha$:
\be
S=-2\pi a\frac{\partial}{\partial\alpha}F,
\ee
so we see that the terms in $F^{\rm TM}_{(1)}$ linear in $\alpha$ and inverse
linear in $\alpha$ violate the Nernst heat theorem,
and therefore seem unphysical.  So, there is motivation
for simply omitting those terms.  
We might think that it is the dimensional quantities $\tau$ and $a\delta$ which
are the fixed regulators, which would suggest a scaling argument for removing
these terms as irrelevant, but this may 
be  incorrect, since a direct calculation
\cite{prachi}
of the stress tensor shows that the principle of virtual work, requiring that
the stress on the sphere be the negative derivative of the free energy with
respect to $a$, shows that the quantities $\tau$ and $\delta$ 
must be regarded as
constant, so that terms in Eq.~(\ref{totalF1}) depending on $\tilde \tau$
possess $a$ dependence.
However we shall subsequently see two further
 justifications for ``renormalizing'' these terms away. The infrared
and ultraviolet divergent terms encountered here are quite different from
the logarithmic terms seen in Eqs.~(\ref{n=0sc}) and (\ref{dfl2}).  The present
divergent terms are power divergences, and so are more convincingly 
removed.
Adopting such a prescription leaves only
the finite terms in the free energy and the entropy
\be
\hat F^{\rm TM}_{(1)}=\frac{\lambda_0}{4\pi a}\left(
\ln\frac{2\sinh\alpha}\alpha
+\frac{\alpha^2}{18}\right),\quad \hat S^{\rm TM}_{(1)}=-\frac{\lambda_0}2
\left(\coth\alpha-\frac1\alpha+\frac\alpha9\right).\label{hatFS}
\ee
Unfortunately, this TM contribution to the entropy, like the TE contribution
(\ref{ste1}),
is always negative, and the sum of the two is linear in the temperature:
\be
\hat S_{(1)}=S^{\rm TE}_{(1)}+\hat S^{\rm TM}_{(1)}=-\frac29 \lambda_0\alpha.
\ee
The first-order renormalized  entropy is displayed in Fig.~\ref{figSx}.
Both contributions to the entropy are now negative.

The same result (\ref{totalF1}) is obtained if instead of direct 
summation, the Euler-Maclaurin formula (around $n=1$, because the
summand is not analytic
at the origin) is used.  However, remarkably, the Abel-Plana formula gives
directly the finite part with all the divergences confined to the temperature-%
independent part.  This is actually not surprising, because analytic 
regularization techniques omit power divergences.
The Abel-Plana formula
\be
\sum_{n=0}^\infty{}'f(n)=\int_0^\infty dt\,f(t)
+i\int_0^\infty dt\frac{f(it)-f(-it)}
{e^{2\pi t}-1},\label{ap}
\ee
concentrates all the divergent terms in the first temperature-independent 
integral.  All the temperature dependence is contained in
\be
\Delta F^{\rm TM}_{(1)}=-i\lambda_0T\int_0^\infty dn\frac{\cos n\alpha \tilde
\tau}{e^{2n\pi}-1}\frac1{n\alpha}\left[-e_0'(i n\alpha)s_0'(i n\alpha)+
\frac\partial{\partial x}\frac\partial{\partial y}\frac{xy}{X}e^{-iX}+(c.c.)
\right]_{x=y=n\alpha},
\ee 
where $X=\sqrt{x^2+y^2-2xy\cos\delta}$, which is just the Minkowski version
of the Euclidean form in Eq.~(\ref{Fab}).
Now, because we have sines and cosines instead of real exponentials, 
contributions are finite, and we can set $\tau\to0$:
\be
\Delta F^{\rm TM}_{(1)}=\frac{\lambda_0}{\pi a}\int_0^\infty \frac{dx}x
\frac1{e^{x/aT}-1}\left(1-\cos^2 x+\frac{x^2}{3}\right).
\ee
This may be easily shown, for example by Borel summation,  to yield precisely 
the free energy and entropy shown in Eq.~(\ref{hatFS}).

\section{Low-temperature behavior}
\label{sec:lowt}
In this section we consider arbitrary finite coupling $\lambda_0$, but examine
the behavior of the free energy for low temperature.  Such can be readily
extracted by use of the Euler-Maclaurin formula (\ref{em}), which will 
concentrate all the ultraviolet divergences in the temperature independent part
of the free energy.  The integral term in Eq.~(\ref{em}), as noted before, 
contributes only to the temperature-independent part.

\subsection{TE low temperature behavior}
\label{sec:telow}
As noted in Sec.~\ref{sec4} the lowest-order appearance of an odd term in
$\zeta_n$ occurs for $l=1$:
\be
\frac{e_1(x)s_1(x)}x\sim \frac13-\frac2{15} x^2+\frac19 x^3+\dots,\quad x\ll 1.
\label{e1s1}
\ee 
The leading low-temperature correction comes by expanding the logarithm in 
powers of $\lambda_0$.  The $k$th order term arising from the Bernoulli
series in the Euler-Maclaurin formula is
\be
\Delta F^{{\rm TE},T\to0}_{(k)}
=-T\frac{B_4}{4!}(2\pi T)^3\left(\frac\partial{\partial x}
\right)^3 3
(-1)^{k+1}a^3\lambda_0^k\left(\frac13\right)^{k-1}\frac{x^3}9\bigg|_{x=0}=
-a^3\left(-\frac{\lambda_0}3\right)^k\frac{\pi^3 T^4}{15}.
\ee
When this is summed over all $k$ we get
\be
\Delta F^{{\rm TE},T\to0}(\lambda_0)=\frac{(\pi a)^3}{15}T^4
\frac1{1+3/\lambda_0}\to \frac{(\pi a)^3}{15}T^4,
\label{nonpertlowtTE}
\ee
where the last replacement is the strong-coupling limit (which may be directly
confirmed). 
The corresponding entropy is
\be
S^{{\rm TE},T\to 0}(\lambda_0)=-\frac{\alpha^3}{30}\frac1{1+3/\lambda_0},
\label{lowTarbl}
\ee
which for $O(\lambda_0)$ coincides with the result found at the end of
Sec.~\ref{sec:telambda}, and in $O(\lambda_0^2)$ agrees with the entropy
computed from the $O(\alpha^4)$ term in Eq.~(\ref{dfl2}), which constitutes
further evidence of the irrelevance of the infrared-sensitive logarithmic
term there. 
 Comparing with the total
low-temperature correction (\ref{lowtsc}), we see that the strong-coupling
 limit of the TM contribution must be
\be
\Delta F^{{\rm TM},T\to0}_\infty=-\frac2{15}(\pi a)^3 T^4.\label{TMltsc}
\ee
The corresponding entropies, $S=-\partial F/\partial T$, are negative for
the TE contribution, and positive for the TM.  As for the plate \cite{ly}, the
latter overwhelm the former.

It might well be objected that since the summand is not analytic
at $n=0$, it would be better to apply the Euler-Maclaurin formula about $n=1$,
as in Eq.~(\ref{emn1}). Doing so in this case yields exactly the same result as
 found here, but in the next subsection, we will see that expansion
about $n=1$ is essential to
get the result in the TM case, where the singularity at the origin is more
severe.

These strong-coupling 
results were given by Bordag et al.~\cite[Sec.~9.5.1]{bkmm}, however 
with an additional $T^3$ term in the free energy subtracted in the case
of the TE contribution, and the same term added to the TM contribution, so
that the total contribution remains unchanged.
This followed upon the earlier suggestion by Geyer et al.~\cite{geyer},
who considered rectangular boxes.  These subtractions, motivated
by the heat-kernel analysis of the Weyl expansion, were justified by
requiring, perhaps dubiously,
 that such a term not be present at high temperature.  We will see
exactly that $T^3$ term when we study the 
high-temperature strong-coupling limit
in Sec.~\ref{sec:sctetm}.  
It seems not possible to make such a subtraction here
because we are considering arbitrary coupling, where the geometrical 
considerations applied in Ref.~\cite{bkmm} must be generalized \cite{vass}.

\subsection{TM low temperature behavior}
\label{sec:tmlow}

For the TM free energy we write
\be
F^{\rm TM}=T\sum_{n=0}^\infty{}'g(n),\quad g(n)=\cos n\alpha\tilde\tau
\sum_{l=1}^\infty (2l+1)P_l(\cos\delta)\ln\left(1-\frac{\lambda_0}x
e'_l(x)s_l'(x)\right).
\ee
For small $x$, the quantity in the logarithm is singular,
\be
\frac{e_l'(x)s_l'(x)}x\sim -\frac{l(l+1)}{(2l+1)x^2}-\frac{3+2l(l+1)}{(4l^2-1)
(2l+3)}+O(x^2 \,\,\mbox{or}\,\,x^{2l-1}),\quad x\ll1. \label{smargtm}
\ee
The special role of $l=1$ is evident.  To define the $n=0$ term
as before  we introduce the photon mass parameter $p$,
so up to terms that vanish with $p$,
\be
\frac12 g(0)=\frac12\sum_{l=1}^\infty (2l+1)P_l(\cos\delta)\ln\frac{\lambda_0
l(l+1)}{(2l+1)p^2}.
\ee
Because $\alpha$ is also very small, we have the leading term
\be
\frac12g(1)\approx
\frac12\sum_{l=1}^\infty (2l+1)P_l(\cos\delta)\ln\frac{\lambda_0
l(l+1)}{(2l+1)\alpha^2},
\ee
and the integral term gives
\be
-\int_0^1 dn \, g(n)=-\sum_{l=1}^\infty (2l+1)P_l(\cos\delta)
\left[\ln\frac{\lambda_0l(l+1)}{(2l+1)\alpha^2}+2\right].
\ee
Adding to these 
the contribution of the first term in the Bernoulli series gives
\be
F^{\rm TM}_{O(T)}\sim T\ln\frac{2\pi p}\alpha.
\ee
This term is, of course, identical to  the $T\ln T$ term 
we found in strong coupling
in Eq.~(\ref{n=0sc}), and should be omitted for the same reasons.

It is easy to check that the terms of $O(T^3)$ coming from the integral,
the $\frac12g(1)$ term, and the first Bernoulli term all cancel.  
The surviving $T^4$
behavior again receives canceling contributions from these three places,
but arises entirely from the exceptional $l=1$ term,
where 
\be
\ln\left(1-\lambda_0\frac{e_1'(x)s_1'(x)}x \right)\sim \ln\frac{2\lambda_0}
{3x^2}+x^2\left(\frac3{2\lambda_0}+\frac7{10}\right)-\frac23x^3+\dots.
\ee
Thus the leading contribution to the TM free energy is
\be
F^{\rm TM}_{O(T^4)}
\sim-\frac23T\alpha^3 3\left(-\frac{B_4}{4!}6\right)=-\frac2{15}
(\pi a)^3 T^4,\label{sctm}
\ee
just as stated above, in Eq.~(\ref{TMltsc}).

So we have recovered the strong-coupling limit.  Still in the low-temperature
context, we can develop an expansion in 
$\xi=\frac{\alpha}{\sqrt{2\lambda_0/3}}$, which we regard as 
arbitrary.\footnote{We recall that in the $\delta$-function plate,
we developed the TM free energy in terms of a strong-coupling expansion,
as a series in $T/\lambda_0$ \cite{ly}.}  This
arises again entirely from the $l=1$ term in the angular momentum expansion.
(Higher terms in $l$ will yield only higher-order terms in $T$.)
This is achieved by expanding 
\be
\ln\left[1+\left(\frac3{2\lambda_0}+\frac7{10}\right)x^2-\frac23x^3\right]
=\sum_{p=1}^\infty(-1)^{p-1}\frac1p\left[\left(\frac3{2\lambda_0}+\frac7{10}
\right)x^2-\frac23x^3\right]^p.
\ee
This leads to the Bernoulli term (about $n=1$)
\be
F^{\rm TM}_B=-3T\sum_{k=1}^\infty \frac{B_{2k}}{(2k)!}\alpha^{2k-1}\left(
\frac{\partial}{\partial\alpha}\right)^{2k-1}\sum_{p=1}^\infty\frac{(-1)^{p-1}}
{p}\sum_{r=0}^p\left(\begin{array}{c}p\\r\end{array}\right)\left(\frac{3}{2
\lambda_0}+\frac7{10}\right)^{p-r}\left(-\frac23\right)^r\alpha^{2p+r},
\ee
where the derivative amounts to inserting the factor $(2p+r)!/(2p+r-2k+1)!$.
Because only $r=1$ can result in $\alpha^3$ dependence (with the 
remaining $\alpha$'s absorbed in the definition of $\xi$), we are led to
\be
F^{\rm TM}_B=-\frac{4\lambda_0^2}{9\pi a}\sum_{k=1}^\infty\frac{B_{2k}}{(2k)!}
\sum_{p=1}^\infty(-1)^p\frac{(2p+1)!}{(2p-2k+2)!}\xi^{2p+2}.
\ee
The sum on $p$ may be readily carried out:
\begin{subequations}
\bea
k=1: \quad&&\sum_{p=1}^\infty (-1)^p\frac{(2p+1)!}{(2p)!}\xi^{2p+2}
=-\xi^4\frac{3+\xi^2}{(1+\xi^2)^2},\\
k>1:\quad &&\sum_{p=1}^\infty (-1)^p\frac{(2p+1)!}{(2p-2k+2)!}\xi^{2p+2}=
(-1)^{k+1}(2k-1)!\xi^{2k}(1+\xi^2)^{-k}\cos(2k\arctan\xi).
\eea
\end{subequations}
Now we notice that
\be
\frac\xi{\sqrt{1+\xi^2}}e^{i\arctan\xi}=\frac\xi{1-i\xi}, 
\ee
so then we have
\be
F^{\rm TM}_B=\frac{4\lambda_0^2}{9\pi a}\left\{\frac{B_2}2\frac{\xi^4(3+\xi^2)}
{(1+\xi^2)^2}+\sum_{k=2}^\infty (-1)^k\frac{B_{2k}}{2k}\frac12\left[
\frac{\xi^{2k}}{(1+i\xi)^{2k}}+\frac{\xi^{2k}}{(1-i \xi)^{2k}}\right]\right\}.
\ee
Finally,  the sum on $k$ may be recast as an integral,
\be
F^{{\rm TM},T\to0}(\xi)
=\frac{4\lambda_0^2}{9\pi a}\left\{\frac{B_2}2\frac{\xi^4(3+\xi^2)}
{(1+\xi^2)^2}+\Re\int_0^\infty \frac{dt}t e^{-t}\left[-1+\frac1{12} x^2t^2+
\frac{xt}2\cot\frac{xt}2\right]-\frac{\xi^4}{2(1+\xi^2)}
+\frac12[\xi^2-\ln(1+\xi^2)]\right\},\label{lowtmexact1}
\ee
where
\be
x=\frac\xi{1+i\xi},\quad \xi=\frac\alpha{\sqrt{2\lambda_0/3}},
\ee
and where we have now added in as  the last two terms
 the contributions from the 
$\frac12g(1)$ term and the integral term in the Euler-Maclaurin formula,
respectively. (The latter is worked out similarly to the way we computed
the derivatives in the Bernoulli sum.)  For strong coupling
\be
F^{{\rm TM},T\to0}(\xi)\sim -\frac{4\lambda_0^2}{9\pi a}\left(\frac{\xi^4}{120}
+\frac{\xi^6}{252}+\frac{\xi^8}{240}+\dots\right)\sim-
\frac{\alpha^4}{120\pi a},\quad \xi\ll1,
\ee
which reproduces Eq.~(\ref{sctm}), of course. For weak coupling
\be
F^{{\rm TM},T\to0}(\xi)= \frac{4\lambda_0^2}{9\pi a}\left[
\frac{\xi^2}{12}+\gamma-\ln\xi+\sum_{n=1}^\infty
\frac{(-1)^n}{\xi^{2n}}\zeta(2n+1)\right]
\sim\frac{\lambda_0}{18\pi a}\alpha^2+\frac{4\lambda_0^2}{9\pi a}\left(\gamma
-\ln\xi\right),
\quad \xi\gg1,\label{lowtmfe}
\ee
where the first term exactly agrees with 
 the small-temperature limit of that found in Eq.~(\ref{hatFS}),
 without the divergent terms seen in Eq.~(\ref{totalF1}).
We can sum the weak-coupling expansion into a closed form:
\be
F^{{\rm TM},T\to0}(\xi)
=\left(\frac{2\lambda_0}{3}\right)^2
\left[\frac{\xi^2}{12}-\ln\xi-\Re\psi\left(1+\frac{i}\xi\right)\right],\label{lowtmexact2}
\ee
which can be shown numerically to coincide with Eq.~(\ref{lowtmexact1}).

\begin{figure}
\includegraphics{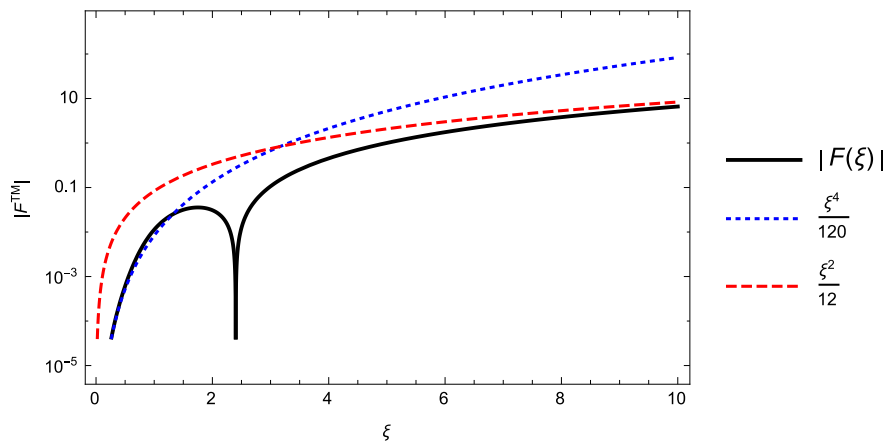}
\caption{\label{lowtm} The absolute value
 of the TM free energy for low temperature,
as a function of the coupling defined through $\xi=\alpha \sqrt{\frac{3}{2
\lambda_0}}$. The overall factor $(2\lambda_0/3)^2/\pi a$ has been pulled out.
 The dotted and dashed lines are the large and small coupling
limits, respectively. For strong coupling, the slope is negative, and hence
the entropy is positive, but for sufficiently weak coupling, the entropy
changes sign.  The cusp indicates where the free energy changes sign.}
\end{figure}

Two lessons are thus learned: That the subtraction (renormalization)
procedure leading to 
the perturbative free energy and entropy (\ref{hatFS}) apparently is valid, 
and that the free energy develops a positive  slope (the entropy becomes
negative) for large enough $\xi$, small enough coupling.  
The low-temperature TM free energy is plotted in 
Fig.~\ref{lowtm} as a function of $\xi$.  Now it is seen that the
free energy starts with  negative slope as a function of temperature, for
large coupling, but at $\xi=1.75271$ the sign of the slope changes, so the 
corresponding entropy turns negative.  The change occurs for $\lambda_0=
0.488282 \alpha^2$.

\section{High-temperature limit}
\label{sec:hit}
Finally, we return to high temperature, both for finite and large coupling,
where unlike in Sec.~\ref{sec4} we consider the TE and TM contributions separately.
These limits are captured completely by the uniform asymptotic expansion
for the Bessel functions, in contrast to the low-temperature limit, and
because the structures encountered in the sums are quadratic forms, it is 
particularly convenient to abandon the point-splitting regularization adopted
heretofore, and use analytic regularization and the generalized Chowla-Selberg 
asymptotic formulas.
In contrast, point splitting yields formulas that are rather complicated
to evaluate.  We start by examining the TE contribution. 
\subsection{TE contribution}
The high-temperature behavior is captured from
the uniform asymptotic expansion, which can be written,
 as in Eq.~(\ref{uaelog})
\be
\ln\left(1+\frac{\lambda_0}x e_l(x)s_l(x)\right)\sim 
\sum_{k=1}^\infty \frac{a_{\rm TE}^{(k)}(t)}{(2\nu)^k},\quad \nu\gg1,
\label{uaelogte}
\ee
where the first four expansion coefficients are
\begin{subequations}
\bea
a_{\rm TE}^{(1)}(t)&=&\lambda_0 t,\label{ate1}\\
a_{\rm TE}^{(2)}(t)&=&-\frac12\lambda_0^2 t^2,\\
a_{\rm TE}^{(3)}(t)&=&\lambda_0\frac{t^3}2(1-6t^2+5t^4)+\lambda_0^3
\frac{t^3}3,\\
a_{\rm TE}^{(4)}(t)&=&-\frac{\lambda_0^2t^4}{2}(1-6t^2+5t^4)
-\frac{\lambda_0^4 t^4}4.
\eea
\end{subequations}

It seems the most effective way to extract the high-temperature dependence
is through use of the Chowla-Selberg formula as generalized by Elizalde
\cite{elizalde1,elizalde2}.  That is, we will discard the point-split 
regularization we have used hitherto, and use the formula
\bea
&&\sum_{n=0}^\infty\sum_{l=0}^\infty \frac1{[c(l+b)^2+d\alpha^2(n+a)^2]^{s}}
\equiv E_2(s;c,d\alpha^2;b,a)\nn\\
&=&\frac{(d\alpha^2)^{-s}}
{\Gamma(s)}\sum_{m=0}^\infty \frac{(-1)^m}{m!}\Gamma(s+m)
\left(\frac{c}{d\alpha^2}\right)^m\zeta(-2m,b)\zeta(2s+2m,a)\nn\\
&&\quad\mbox{}+\frac{(d\alpha^2)^{-s}}2\left(\frac{\pi d\alpha^2}{c}
\right)^{1/2}\frac{\Gamma(s-1/2)}{\Gamma(s)}\zeta(2s-1,a)\nn\\
&&\quad\mbox{}+\frac{2\pi^s}{\Gamma(s)}\cos(2\pi b) c^{-s/2-1/4}
(d\alpha^2)^{-s/2+1/4}\sum_{n=1}^\infty \sum_{m=0}^\infty n^{s-1/2}
(m+a)^{-s+1/2}
K_{s-1/2}\left(2\pi\sqrt{\frac{d\alpha^2}{c}}n(m+a)\right).\label{cs}
\eea
Here, the various parameters, $a$, $b$, $c$, and $d$ are introduced so that 
the desired structures can be obtained by appropriate differentiation.  
Afterwards, we set $a=1$, $b=3/2$, $c=d=1$. We shift the $n>0$ and $l>0$ sums
so they start at zero.
 Because to the prefactor of 
$\nu$ in the summand for the free energy, there is always one derivative with
respect to $b$, so the second term above does not contribute.  Neither does the
last term, because $b$ is an integer plus one-half.  Thus, only the first term 
in Eq.~(\ref{cs}) survives.

\subsubsection{High-temperature, fixed coupling}
Let us start with the first term in the uniform asymptotic expansion 
(\ref{ate1}). Note that the leading term in the uniform
asymptotic expansion is exact for $n=0$.
By $\zeta$-function regularization, the $n=0$ term yields
\be
F^{(1){\rm TE}}_{n=0}
=\frac{\lambda_0 T}2\zeta(0,3/2)=-\frac{\lambda_0 \alpha}{4\pi a}.
\ee
while the remainder of the $n$ sum gives
\be F^{(1){\rm TE},T\to\infty}_{n>0}
=-\lambda_0T\frac1{2(s-1)}\frac{\partial}{\partial b}E_2(
s-1;1,\alpha^2;b,1)\bigg|_{s\to1/2}.
\ee
The $m=0$ term in the Chowla-Selberg formula (\ref{cs})
now gives immediately
\be
F^{(1){\rm TE},T\to\infty}_{n>0,m=0}=\frac{\lambda_0\alpha^2}{24\pi a},
\label{leadinghit}
\ee 
while the $m=1$ term gives
\be
F^{(1){\rm TE},T\to\infty}_{n>0,m=1}
=-\frac{\lambda_0}{4\pi a}\frac{11}{12}\left(\gamma-\ln\alpha+
\frac1{2s-1}\right),
\ee
where the divergent term is irrelevant, since it is $T$-independent.  In this
way, we find
\be
\Delta F^{(1){\rm TE},T\to\infty}
=\frac{\lambda_0}{4\pi a}\left(\frac{\alpha^2}6-\alpha+\frac{11}
{12}\ln\alpha\right),
\ee
which is nearly that expected from the exact $O(\lambda_0)$ result 
(\ref{te1exact}):
\be
T\to\infty:\quad \Delta F^{(1){\rm TE}}
\sim\frac{\lambda_0}{4\pi a}\left(\frac{\alpha^2}6-
\alpha+\ln\alpha\right).\label{te1exactht}
\ee
The missing contribution to the logarithmic term comes form
the $O(\lambda_0)$ part of $a^{(3)}_{\rm TE}$.  Using the Chowla-Selberg
formula again, but now also differentiating with respect to $c$ and $d$ as
well, we find
\be
F^{(3){\rm TE},T\to\infty}_{O(\lambda_0)}
\frac{\lambda_0}{4\pi a}\frac1{12}\ln\alpha.\label{a3cont}
\ee
There are no higher power temperature
 corrections in $O(\lambda_0)$ (beyond that displayed, the temperature
 corrections
are exponentially small).  Indeed, when the $m=2$ contribution from the lowest
order UAE is added to the $m=1$ contribution
 from the second order UAE, and the $m=0$
contribution from the third order UAE (not displayed in the above formulas),
 which would potentially give a 
contribution to the free energy of order $\alpha^{-2}$, we obtain zero.

Next, we look at the  $T\to \infty$ contribution coming from $a^{(2)}$:
\be F^{(2){\rm TE},T\to\infty}
= -\frac{\lambda_0^2 T}4\sum_{n=0}^\infty{}'
\sum_{l=1}^\infty \frac\nu{(\nu^2+n^2\alpha^2)^s},\quad s\to1, 
\ee
where the $n=0$ contribution is
\be
F^{(2){\rm TE},T\to\infty}_{n=0}
=-\frac{\lambda_0^2 \alpha}{16\pi a}\left[-2^{2s-1}+\left(2^{2s-1}-1
\right)\zeta(2s-1)\right],
\ee which is divergent as $s\to 1$.  This divergence is canceled by
another divergence occurring in the
remaining contributions,
\be
F^{(2){\rm TE},T\to\infty}_{n>0}
=-\frac{\lambda_0^2 T}8 \frac{\alpha^{2-2s}}{1-s}
\zeta(2s-2)+\dots.
\ee
When these are added together, we obtain a finite result:
\be  F^{(2){\rm TE},T\to\infty}
=-\frac{\lambda_0^2 T}8 \left(-2+\gamma+\ln\frac{2\alpha}\pi
-\frac{11}{12}\zeta(2)\frac1{\alpha^2}+
O(\alpha^{-4})\right).\label{eff2}
\ee
Here we have kept a subdominant term that will not contribute in the
high-temperature, fixed $\lambda_0$ regime, because we will see this is
important in strong coupling.

All that remains are the higher order terms in $\lambda_ot$, which 
for fixed $\lambda_0$ only
contribute at $n=0$. (Again the UAE is exact in this case.)
These give a contribution linear in $T$ which is easily
summed:
\bea
F^{(>2){\rm TE},T\to\infty}_{n=0}
&=&T\sum_{l=1}^\infty \nu\left[\ln\left(1+\frac{\lambda_0}{2\nu}\right)-
\frac{\lambda_0}{2\nu}+\frac{\lambda_0^2}{8\nu^2}\right]
=T\bigg[-\frac1{24}-\frac{13}{24}\ln2+\frac12\ln G+\frac{\lambda_0}2
-\frac{\lambda_0}{4}\ln2\pi\nn\\
&&\quad\mbox{}-\frac{\lambda_0^2}8(1-\gamma-2\ln 2)
+\frac{\lambda_0}2\ln\Gamma\left(\frac{3+\lambda_0}{2}\right)-\zeta^{(1,0)}
\left(-1,\frac{3+\lambda_0}2\right)\bigg].\label{hthl3}
\eea
Adding Eqs.~(\ref{te1exactht}), (\ref{eff2}), and (\ref{hthl3}) together,
we find
\bea
F^{{\rm TE},T\to\infty}(\lambda_0)
&\sim&\frac1{2\pi a}\bigg\{\alpha\left[-\frac1{24}
-\frac{13}{24}\ln2+\frac12\ln G\right]+\lambda_0\left[\frac{\alpha^2}{12}
+\frac12\ln\alpha-\frac\alpha4\ln 2\pi\right]
+\frac18\lambda_0^2\alpha\left(1-\ln\frac\alpha{2\pi}\right)\nn\\
&&\quad\mbox{}+\frac12\lambda_0\alpha\ln\Gamma\left(\frac{3+\lambda_0}2\right)
-\alpha\zeta^{(1,0)}\left(-1,\frac{3+\lambda_0}2\right)
+\frac{11}{96}\frac{\lambda_0^2}{\alpha}\zeta(2)\bigg\}.\label{fixllargeT}
\eea
The leading term for high temperature, fixed coupling, is just that given in
Eq.~(\ref{leadinghit}):
\be
F^{{\rm TE},T\to\infty}(\lambda_0)\sim \frac{\lambda_0}{24\pi a}\alpha^2,
\quad aT\gg1.
\ee

Particularly interesting here is the strong-coupling limit, $\lambda_0\to 
\infty$:  
\bea
F^{{\rm TE},T\to\infty}_{\lambda_0\to\infty}
&\sim& \frac1{2\pi a}\left[
-\frac{\lambda_0^2\alpha}{16}\left(1+
2\ln\frac{\alpha}{\lambda_0 \pi}\right)+\frac{\lambda_0}{12}(\alpha^2
+6\ln\alpha)-\frac{\alpha}{24}(1+2\ln2-12\ln G+11\ln\lambda_0)
+\frac{11}{12}\frac{\lambda_0^2}\alpha \frac{\pi^2}{48}\right],\nn\\
&&\qquad aT\gg\lambda_0\gg1.\label{sc0}
\eea
Although the $\lambda_0 \alpha$ term canceled, the high-temperature
TE free energy does not possess a strong-coupling limit independent of
$\lambda_0$. This is because this result is valid for $aT\gg\lambda_0\gg1.$
That is, we are taking the high-temperature limit before we pass to strong
coupling.  To reverse the order of limits, we have to consider additional
terms.

\subsubsection{Strong-coupling limit}
\label{sec:scl}
To get the entropy in the limit $\lambda_0\gg aT\gg 1$ we have to add the
leading correction, because we only included the $n=0$ term in 
Eq.~(\ref{hthl3}):
\be
F^{\rm TE}_{\rm corr1}
=T\sum_{n,l=1}^\infty 2\nu\sum_{k=3}^\infty 
\frac{(-1)^{k-1}}k\left(\frac{\lambda_0}{2\nu}\right)^kt^k
=\frac1{2\pi a}\left[-\frac{\lambda_0^3}{12}\ln\alpha+
2\alpha^3f(\lambda/2\alpha)-\frac{11}{12}\alpha g(\lambda/2\alpha))\right],
\label{sccorr1}
\ee
where the first term arises by taking a limit in the $k=3$ term, 
and the sum
over the remaining $k$ sum yields
\bea
f(x)&=&-\frac{1}{24}\left[x-3x^2-6x^3+8\gamma x^3+6 x^2\ln 2\pi
-\frac{3\zeta(3)}{\pi^2}-12\zeta^{(1,0)}(-2,1+x)
+24x\zeta^{(1,0)}(-1,1+x)\right]
\nn\\
&\to& \frac19x^3(4-3\gamma-3\ln x)+\frac18x^2(1-2\ln 2\pi x)-\frac1{12}x+
\frac{\zeta(3)}{8\pi^2}+\dots,
\eea
and
\be
g(x)=\frac{\pi^2}{12}x^2-\gamma x-\ln\Gamma(1+x)
\to\frac{\pi^2}{12}x^2+(1-\gamma-\ln x)x-\frac12\ln2\pi x,
\ee
where the last forms correspond to the strong-coupling limit, $x=
\frac{\lambda_0}{2\alpha}\gg1$.  When these are substituted into 
Eq.~(\ref{sccorr1}), the $\lambda_0^3\ln\alpha$ term cancels, and the
leading term in $g(x)$ cancels the subleading term in Eq.~(\ref{eff2}),
\be
\Delta F^{{\rm TE},T\to\infty}_{\rm corr1}
=\frac1{2\pi a}\left[-\frac{\lambda_0^2}{\alpha}\frac{11}{12}\frac{\pi^2}{48}
+\alpha\frac{\lambda_0^2}{16}
\left(1-2\ln\frac{\pi\lambda_0}\alpha\right)
-\alpha^2\frac{\lambda_0}{12}+\alpha^3\frac{\zeta(3)}{4\pi^2}
+\frac{11}{24}(\lambda_0+\alpha)\ln\frac{\pi\lambda_0}{\alpha}+\dots\right],
\label{sccorr1f}
\ee
where the omitted terms are either independent of $T$ or 
 go to zero as $\lambda_0\to \infty$.

We also have to keep the next term in the UAE, which gives
\be
F^{{\rm TE},T\to\infty}_{\rm corr2}
=\frac{T}2\sum_{n,l=1}^\infty2\nu\sum_{k=2}^\infty (-1)^{k-1}\lambda_0^k
\left(\frac{t}{2\nu}\right)^{k+2}(1-6t^2+5t^4)
\to\frac3{64\pi a}\alpha-\frac{\lambda_0}{48\pi a}\ln\alpha,\label{sccorr2}
\ee
where again the strong-coupling limit is displayed.  Note that the last term
here cancels the contribution (\ref{a3cont}).

When these corrections (\ref{sccorr1f}) and (\ref{sccorr2}) are added to
previous result
(\ref{sc0}), we
 recover the expected perfect-conductor limit for the
TE contribution given below: 
\be
F^{{\rm TE},T\to\infty}_{\infty}=\frac1{2\pi a}\left[\alpha^3\frac{\zeta(3)}
{4\pi^2}-\frac{11}{24}\alpha\ln\frac{\alpha}{2\pi}
-\frac{13}{24}\alpha\ln2+\frac12\alpha\ln G+\frac5{96}\alpha\right].
\label{tesclimit}
\ee

In Ref.~\cite[Sec.~9.5.2]{bkmm} the free energy for a Dirichlet spherical
shell is given.  If the $l=0$ term is subtracted from that, 
$F^{\rm{TE},D}_{l=0}=(T/2)\ln 2aT$, this coincides with our result, except
for the leading $\alpha^3$ term.  This is precisely the term that is
subtracted off by the procedure advocated in that reference, as mentioned 
already at the end of Sec.~\ref{sec:telow}.  The objection to this term is
that it grows as the area of the sphere; however, such a subtraction does
not seem possible in our general analysis, in which the perfect conductor
boundary conditions are only obtained as a limit.

\subsection{TM Contribution}
\label{TMstrong}
For $n\ne0$, the UAE again
should capture the leading high-temperature contribution.
As in Eq.~(\ref{uaelogte}),
\be
\ln\left(1-\frac{\lambda_0}x e_l'(x)s_l'(x)\right)\sim 
\sum_{k=1}^\infty \frac{a_{\rm TM}^{(k)}(t)}{(2\nu)^k},\quad \nu\gg1,
\label{uaelogtm}
\ee
where the first four expansion coefficients are
\begin{subequations}\label{uaetm}
\bea
a_{\rm TM}^{(1)}(t)&=&\frac{\lambda_0}{z^2 t},\label{atm1}\\
a_{\rm TM}^{(2)}(t)&=&-\frac{\lambda_0^2}{2z^4 t^2},\\
a_{\rm TM}^{(3)}(t)&=&-\frac{\lambda_0}{2z^2}t(1-6t^2+7t^4)+\frac{\lambda_0^3}
{3 z^6t^3},\\
a_{\rm TM}^{(4)}(t)&=&\frac{\lambda_0^2}{2z^4}(1-6t^2+7t^4)
-\frac{\lambda_0^4}{4 z^8 t^4}.
\eea
\end{subequations}

We will here content ourselves with the 
 consideration of  the $O(\lambda_0)$ contribution.
For $n=0$ we have to use the small-argument expansion (\ref{smargtm}) 
regularized by putting in a small photon mass $p$ as before.  
To lowest order in $\lambda_0$,
the part sensitive to $p$ vanishes using zeta-function regularization:
\be
F^{\rm TM}_{n=0,p}=-\frac{\lambda_0 T}{2p^2}\sum_{l=1}^\infty l(l+1)=
-\frac{\lambda_0 T}{8 p^2}[(1-2^2)\zeta(-2)-(1-2^0)\zeta(0)]=0.
\ee
This leaves only the finite remainder
\be
F^{{\rm TM},T\to\infty}_{n=0}=\frac{\lambda_0 T}{4}\sum_{l=1}^\infty\left[
\frac9{(2l+3)(2l-1)}+1\right]=\frac{\lambda_0 \alpha}{4\pi a},
\ee
again using zeta-function regularization.  This is the linear-in-$T$ 
behavior expected from the exact result (\ref{hatFS}).

For $n>0$ we use the UAE (\ref{uaetm}) which gives the leading term
\be
F^{{\rm TM},T\to \infty}_{(1),n>0}
=\lambda_0T 
\sum_{n=0}^\infty\sum_{l=0}^\infty \frac\nu
{(n+1)^2\alpha^2} t^{-2s}, 
\quad \nu=l+b,\quad s\to1/2,
\ee
$t^{-2}$ being the quadratic form seen in Eq.~(\ref{cs}),
where this may be evaluated by differentiating with respect to $d$, and
then using the $m=0$ term in the Chowla-Selberg formula:
\be
F^{(1){\rm TM},T\to\infty}_{m=0}
=\frac{\lambda_0\alpha^2}{72\pi a},
\ee
which is exactly the expected quadratic $T$ dependence seen in 
Eq.~(\ref{hatFS}).

The $m=1$ term here gives in the same way a contribution to the logarithm
term:
\be
F^{(1){\rm TM},T\to\infty}_{m=1}
=\frac{\lambda_0}{4\pi a}\frac{11}{12}\ln\alpha.
\label{firstlog}
\ee
However, there is a similar contribution coming from the 
order $O(\lambda_0)$ part of the $a_{\rm TM}^{(3)}$ 
term
\be
F^{(3){\rm TM},T\to\infty}_{n>0}
=-\frac{\lambda_0 T}8 
\sum_{n,l=0}^\infty
\frac{\nu}{(n+1)^2\alpha^2}t^{2s}(1-6t^2+7t^4),\quad s\to1/2.
\ee
Again by differentiating with respect to parameters we find,
in the large temperature limit, that only the $m=0$ term in the Chowla-Selberg
formula contributes, yielding
\be
F^{(3){\rm TM},T\to\infty}=-\frac{\lambda_0}{4\pi a}\frac{23}{12}\ln\alpha.
\ee
When this is added to the previously found logarithmic term (\ref{firstlog})
we obtain exactly the expected result from the exact calculation:
\be
F^{{\rm TM},T\to\infty}_{O(\lambda_0)\mbox{log}}
=-\frac{\lambda_0}{4\pi a}\ln\alpha.
\ee
The net result is just as anticipated from Eq.~(\ref{hatFS}),
\be
\Delta F^{\rm TM}_{O(\lambda_0)}\sim\frac{\lambda_0}{4\pi a}\left(\alpha
+\frac{\alpha^2}{18}-\ln\alpha\right).
\ee
The fact we get the same result as in Sec.~\ref{sec:tmweak}, without the
divergent terms seen in Eq.~(\ref{totalF1}), 
is strong evidence that the minimal subtraction scheme there is valid.

\subsection{Strong-coupling TE and TM contributions}
\label{sec:sctetm}
In Sec.~\ref{sec4} we rederived the entropy for a perfectly conducting sphere 
at high temperatures.  Now we want to extract the TE and TM contributions
for such a sphere.  These seem not to have been presented previously,
although the related Dirichlet result was given in 
Ref.~\cite[Sec.~9.5.2]{bkmm}. (See also the discussion at the end of
Sec.~\ref{sec:scl}.)
The TE contribution is given by
\be
F^{{\rm{TE}}}_\infty=\frac{T}2\sum_{n=-\infty}^\infty e^{i\zeta_n\tau}
\sum_{l=1}^\infty(2l+1)P_l(\cos\delta)\ln\frac{s_l(x)e_l(x)}x.
\ee
The zero Matsubara frequency contribution is
\be
F^{\rm{TE}}_{\infty,n=0}=\frac{T}2\sum_{l=1}^\infty (2l+1)P_l(\cos\delta)\ln
\frac1{2l+1},
\ee
where again we note that the UAE is exact in this case.  Proceeding
analytically, we drop the point-splitting, and find
\be
F^{{\rm TE},T\to\infty}_{\infty,n=0}
=-\frac{T}2\frac\partial{\partial\beta}\sum_{l=1}^\infty
(2l+1)^{1+\beta}\bigg|_{\beta=0}
=-\frac{T}2\left[\frac1{12}+\frac{1}{6}\ln2-\ln G\right].\label{ten=0}
\ee
For $n>0$ we can use the Chowla-Selberg formula (\ref{cs}), starting from  
the leading factor in the UAE,
\bea
F^{(1){\rm TE},T\to\infty}_{\infty,n>0}
&=&2T\sum_{n=1}^\infty\sum_{l=1}^\infty \nu \frac\partial{\partial\beta}
(n^2\alpha^2+\nu^2)^{-\beta/2}\bigg|_{\beta=0}-\frac{11}{24}T\ln 2\nn\\
&=&T\left[-\alpha^2\zeta'(-2)-\frac{11}{24}\ln\frac\alpha{2\pi}
-\frac{11}{24}\ln 2\right].
\label{teleading}
\eea
The 2nd-order term in the UAE yields
\be
F^{(2){\rm TE},T\to\infty}_{\infty,n>0}=T\sum_{n,l=1}^\infty 2\nu
\frac{t^2}{8\nu^2}(1-6t^2+5t^4)
=\frac3{32}T+O(T^{-1}).\
\ee
(This result can also be obtained by the method yielding a
sum over hyperbolic functions used in
 Sec.~\ref{sec4}.)
Adding these components together gives
\be
F^{{\rm TE},T\to\infty}_\infty 
=T\left[\frac{\zeta(3)}{4\pi^2}\alpha^2-\frac{11}{24}\ln
\frac\alpha{2\pi}-\frac{13}{24}\ln 2
+\frac5{96}+\frac12\ln G\right],\label{TESC}
\ee
which coincides with the $\lambda_0\gg aT\gg1$ limit in 
Eq.~(\ref{tesclimit}).
Again, recall that the leading $a^2T^3$ term here is subtracted from
the free energy by the method advocated in Ref.~\cite[Sec.~9.5]{bkmm}.
The same term would be added to the TM contribution, so there is no effect
on the total free energy.  If one were only interested in the perfect
conductor limit, such a procedure seems sensible, but here, with the
context of an arbitrarily-coupled semitransparent sphere, it seems 
unreasonable to implement, so we do not make such subtractions.

The TM contribution is again  more subtle.   For $n=0$ we use the small 
argument expansion (\ref{smargtm}), and
replace $x$ by $p$:
\be
F^{{\rm TM},T\to\infty}_{\infty,n=0}=\frac{T}2\sum_{l=0}^\infty (2l+1)
\ln\frac{l(l+1)}{(2l+1)p^2}
=T\left[\frac{11}{12}\ln 2p+\frac{1}{12}\ln2+\frac1{24}-\frac12\ln G\right]
\label{comp1}
+F',
\ee
where the finite term displayed is the negative of that seen 
in Eq.~(\ref{ten=0}), and the remainder is evaluated as follows:
\bea
F'&=&\frac{T}2\sum_{l=1}^\infty (2l+1)\ln\left(1-\frac1{(2l+1)^2}\right)
=-\frac{T}2\sum_{k=1}^\infty\frac1k\frac1{2^{2k-1}}
\sum_{l=1}^\infty \frac1{(l+1/2)^{\beta(2k-1)}}
\nn\\
&=&-T\left[\frac1{8(\beta-1)}+\frac{13}{12}\ln 2-3\ln G\right],\label{comp2}
\eea
where we have regulated the first term in the $k$-sum by introducing an 
analytic parameter
$\beta$ which tends to 1.
(This same sum, regulated by angular point-splitting, was encountered 
previously in Eq.~(\ref{n0sc}).) The singularity here as $\beta\to1$ will be 
canceled by that in  the $n>0$ contributions.  The latter are captured by
the UAE, as usual, which give, for the leading term
\be
F^{(1){\rm TM},T\to\infty}_{\infty,n>0}
=2T\sum_{n,l=1}^\infty \nu\ln\frac{\sqrt{n^2\alpha^2+\nu^2}}{2n^2\alpha^2}
=T\left[\alpha^2\zeta'(-2)-\frac{11}{24}\ln\frac\alpha{2\pi}
-\frac{11}{24}\ln 2\right].\label{comp3-4}
\ee
The second order contribution
contains a pole when evaluated using Chowla-Selberg (\ref{cs}),
\be
F^{(2){\rm TM},T\to\infty}_{\infty,n>0}=
T\sum_{n, l=1}^\infty 2\nu\left[-\frac{t^2}{8\nu^2}(1-6t^2+7t^4)\right]
=T\left[\frac1{8(\beta-1)}-\frac9{32}-\frac14\ln\frac\alpha{2\pi}\right];
\label{comp5}
\ee
the  pole here (having been regulated in precisely the same way)
cancels that in Eq.~(\ref{comp2}).
Adding all these components (\ref{comp1}), (\ref{comp2}), (\ref{comp3-4}), 
and (\ref{comp5}) together gives
\be
F^{{\rm TM},T\to\infty}_{\infty} =T\left[-\alpha^2\frac{\zeta(3)}{4\pi^2}
-\frac{17}{24}\ln
\frac\alpha{2\pi}+\frac{11}{12}\ln p-\frac{23}{96}+\frac52\ln G-\frac{13}{24}
\ln2\right].\label{TMSC}
\ee
The sum of $F^{\rm TE}_\infty$ and $F^{\rm TM}_\infty$, Eqs.~(\ref{TESC}) and
(\ref{TMSC}) should yield $F^{T\to\infty}_{\infty}$ as in
Eq.~(\ref{hitpc}); in fact,
\be
F^{{\rm TE},T\to\infty}_\infty+F^{{\rm TM},T\to\infty}_\infty =
\frac{11}{12}T\ln\frac{2\pi p}\alpha +F_\infty^{T\to\infty}.\label{tm+te=tot}
\ee
The infrared divergent term is exactly the same as that seen in 
Eq.~(\ref{irfe}).  As argued in
Sec.~\ref{sec4}, that term should be subtracted to obtain the physically
meaningful free energy or entropy.

\section{Summary and Discussion}
\label{sec7}
In this paper we have investigated the entropy of a spherical shell, modeled 
beyond that of a perfect conductor, as a $\delta$-function 
(``semitransparent'') shell of radius $a$.  This study was motivated by our 
desire to better understand the mysterious phenomenon of negative entropy.
Earlier, we had considered a similar model of the $\delta$-function plate,
where the TE entropy was always negative while the TM entropy was positive, and
larger in magnitude, but both approached zero in the strong coupling limit
\cite{ly}.  Apparently the only limit previously studied for the shell was
that of a perfect conductor, without any mode decomposition in 
Ref.~\cite{bd,teo}, while the mode decomposition can be 
found in Ref.~\cite{bkmm}.
Here we accomplish a full analysis, for arbitrary
coupling,  which exhibits many surprising features.
We will now give a summary of our findings.

In Sec.~\ref{sec3} we derive the general expressions (\ref{gfe})
for the free energy of
an electromagnetic $\delta$-function sphere, decomposed into the TE and TM
modes.  Cancellations occur between the different mode contributions. In
Sec.~\ref{sec4}, devoted to strong-coupling (perfectly-conducting
spherical shell),  we rederive the total (TE plus TM)
free energy correction for low temperature 
[Eq.(\ref{lowtsc})],
\be \Delta F_\infty^{T\to0}\sim -\frac{(\pi a)^3}{15} T^4,\quad aT\ll1
\ee
(where the $\Delta$ refers to the fact this is the correction to the
zero-temperature Boyer energy \cite{boyer}),
and for high temperature [Eq.~(\ref{htscbd})], 
\be F^{T\to\infty}_\infty \sim -\frac{T}4(\ln aT+0.7686), \quad aT\gg 1.
\label{schight}
\ee
However, to obtain these known results \cite{bd}, we removed a contribution
that was infrared sensitive arising from the leading $\ln\zeta_n^2 a^2$ 
term ($\zeta_n=2\pi n T$ being the Matsubara frequency), 
which seems rather a generalization of a contact term [Eq.~(\ref{n=0sc})],
\be 
F^{(0)}_\infty = T\ln \frac\mu{T},\label{lnx2}
\ee
for point-splitting regularization, while this result is multiplied by 11/12
if analytic regularization is used. Here, $\mu$ is a photon mass, introduced
to define the TM mode at zero frequency.

We then considered the opposite limit, weak coupling.  The TE and TM modes
can be exactly evaluated to first order in the coupling $\lambda_0$.  The TE
self-entropy is [Eq.~(\ref{ste1})] ($\alpha=2\pi a T$)
\be
S^{\rm TE}_{(1)}=-\frac{\lambda_0}2\left(\frac13\alpha-\coth\alpha+\frac1\alpha
\right).
\ee
  However, the TM self-entropy exhibits both infrared and ultraviolet
divergences, which were regulated by point-splitting in the angular and 
temporal directions, and by the introduction again of a photon mass.  If these
divergent terms are simply removed (which is done automatically by analytic
regularization, as through use of the Abel-Plana formula) the result is [Eq.~(\ref{hatFS})]
\be
S_{(1)}^{\rm TM}=-\frac{\lambda_0}2\left(\frac19\alpha+\coth\alpha-\frac1\alpha
\right),
\ee
which is, like the TE contribution {\it always negative}.  Notice that the
total entropy is linear in the temperature in this order, as shown in 
Fig.~\ref{figSx}.

In order $\lambda_0^2$ the TE free energy exhibits an infrared divergence,
similar to that in Eq.~(\ref{lnx2}).  This also must be removed in order
to avoid a violation of the Nernst theorem, although here, doing so leads to
a positive contribution to the entropy.

The low-temperature behavior for finite coupling was next explored.  For the
TE contribution, a simple formula is obtained that interpolates between the
weak- and strong-coupling regimes [Eq.~(\ref{lowTarbl})]:
\be
S^{\rm TE}(\lambda_0)\sim-\frac{\alpha^3}{30}\frac1{1+3/\lambda_0},\quad aT
\ll1,
\ee
which shows no sign of the infrared divergence in $O(\lambda_0^2)$ mentioned
just above.
A more complicated formula is found for the TM part [Eq.~(\ref{lowtmexact2})]:
\be
S^{\rm TM}(\lambda_0)=\frac49\lambda_0^2\left[\frac1{12}\xi^2-\ln\xi-\Re\psi
(1+i/\xi)\right],\quad \xi=\sqrt{\frac{3}{2\lambda_0}}\alpha, \quad\alpha\ll1,
\quad \xi\,\,\mbox{arbitrary}.
\ee
This also yields the known strong- and weak-coupling results.

Finally, we considered the high-temperature limit, for arbitrary coupling,
for both TE and TM contributions.  In this situation, we changed our strategy,
and instead of point-splitting, we regulated the double sum over Matsubara
frequencies and angular momentum by use of the generalized Chowla-Selberg
formula in the form given by Elizalde \cite{elizalde1,elizalde2}, which is
particularly convenient for high temperature, since the uniform asymptotic
expansions of the Bessel functions capture all the essential physics there.
For the TE part,
we obtain a rather complicated formula (\ref{fixllargeT}) for the free energy
for fixed $\lambda_0$ and high temperature, which, unsurprisingly, says that
the dominant high-temperature entropy is that given by the $O(\lambda_0)$ 
contribution.  If we want to take the strong-coupling limit, additional terms
in the UAE must be included.  Because of the increased complexity of the TM
contributions, we only extracted, and reproduced, the $O(\lambda_0)$ entropy,
which process, however, vindicated the minimal 
subtraction procedure used in the exact calculation.  The coda consisted of
 computing, directly in strong coupling, the
TE and TM contributions, which should add up to the result (\ref{schight}).
This they do, except for an extra term [Eq.~(\ref{tm+te=tot})]
\be
\Delta F_\infty^{\rm IR}=\frac{11}{12}T\ln\frac{2\pi p}\alpha,\quad p=\mu a,
\ee
This term is the same as the ``contact term'' (\ref{lnx2}), which
we believe should be removed {\it a priori.}

One might think that these surprising findings are a consequence of our use
of the plasma model to describe the dispersive character of the coupling.
However, if the perhaps more realistic Drude model is used, or even a model
of a bound electron so that a characteristic frequency of the binding is introduced,
the situation is not more satisfactory.  Although these modifications change 
the 
infrared behavior, they do not change the ultraviolet behavior, 
and the appearance of
divergent terms in the entropy, and negative TM and total entropies, remains 
present.

At zero temperature, it is typically argued \cite{Graham:2003ib} that the order
 $\lambda_0$
contribution to the energy
should be discarded, since it can be canceled by a counterterm.  This is 
probably
only possible for the $T=0$ contribution, and not for the temperature 
correction.
It is true that if the $O(\lambda_0)$ term were removed from the 
low-temperature
result (\ref{lowtmexact2}) the TM entropy would become positive in that regime 
for
all coupling.  However, this would wreck the internal consistency of the 
problem,
in particular the passage to the strong coupling limit described in 
Sec.~\ref{TMstrong}.
Therefore, this does not appear to be a viable resolution to our difficulty.

\section{Conclusions}
\label{sec:concl}
So after somewhat elaborate calculations, we have obtained unsettling results.
Contrary to expectations, the TM entropy for a $\delta$-function sphere
fails to be  finite, both in the infrared and  in the ultraviolet.  If these
divergent terms are merely subtracted (which is somewhat justified by the
congruence of the results with analytic calculations based on the Abel-Plana
and the Chowla-Selberg formulas)  we find that the TM entropy and the total
entropy are not necessarily positive.  In fact the total entropy is
positive only  if the coupling is sufficiently strong.  The perfectly conducting 
limit is satisfactory, and overcomes the negative interaction entropy between
a perfectly conducting sphere and a perfectly conducting plate, but apparently
for sufficiently imperfect reflectors, this positivity breaks down.  The
nonmonotonicity of the entropy  means also that the specific heat need not be
positive.  The significance of these surprising thermodynamic findings 
merits further study.
 
\acknowledgments
We thank Steve Fulling for extensive discussions, and Michael Bordag for
comments.
We are grateful to the Norwegian Research Council, project number 250346 for 
support of this research. 
LY thanks the Avenir Foundation and the Carl T. Bush Foundation for support.

\end{document}